**A machine learning algorithm for predicting naturalized flow duration curves at human influenced sites and multiple catchment scales**

Michael J. Friedel[1,2,*], Dave Stewart[3,4] , Xiao Feng Lu[4], Pete Stevenson[4], Helen Manly[4], Tom Dyer[4]

[1] University of Colorado, Denver, Colorado, United States (michael.friedel@ucdenver.edu)
[2] Earthquest Consulting Ltd, Auckland, New Zealand 1025 (mike@earthquestconsulting.com)
[3] RainEffects Ltd, Dunedin, New Zealand
[4] Otago Regional Council, Dunedin, New Zealand

KEY POINTS

- Meta models predict naturalised stochastic flow across discrete exceedance probabilities
- Naturalised flow duration simultaneously predicted at human influenced gauge and reach sites
- Naturalised flow duration simultaneously predicted across 7 Strahler stream orders
- Naturalised flow duration predictions agree with independent observations
- Naturalised flow duration prediction by Meta models outperformed SWAT model simulations

GRAPHICAL ABSTRACT

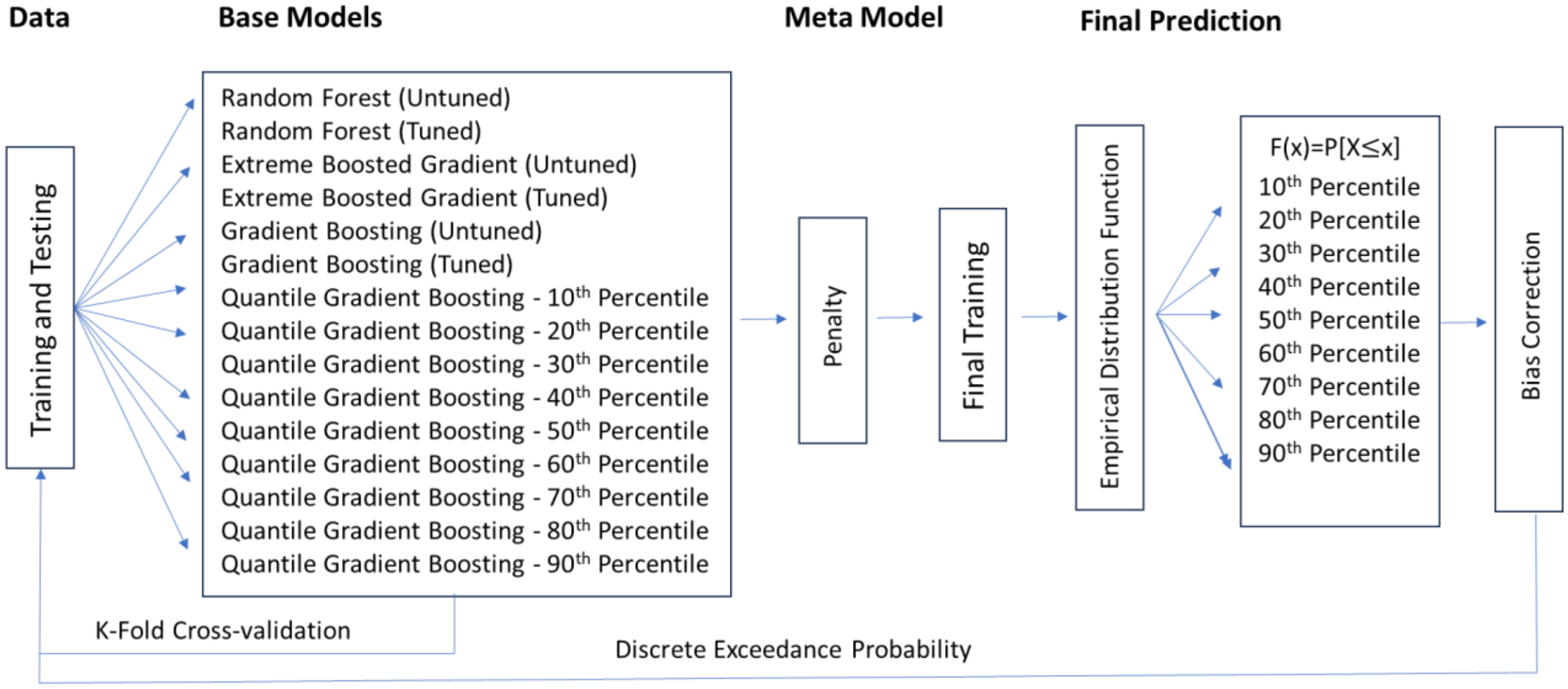

**Abstract**

Regional flow duration curves (FDCs) often reflect streamflow influenced by human activities. We propose a new machine learning algorithm to predict naturalized FDCs at human influenced sites and multiple catchment scales. Separate Meta models are developed to predict probable flow at discrete exceedance probabilities across catchments spanning multiple stream orders. Discrete exceedance flows reflect the stacking of k-fold cross-validated predictions from trained base ensemble machine learning models with and without hyperparameter tuning. The quality of individual base models reflects random stratified shuffling of spilt catchment records for training and testing. A Meta model is formed by retraining minimum variance base models that are bias corrected and used to predict final flows at selected percentiles that quantify uncertainty. Separate Meta models are developed and used to predict naturalised stochastic flows at other discrete exceedance probabilities along the duration curve. Efficacy of the new method is demonstrated for predicting naturalized stochastic FDCs at human influenced gauged catchments and ungauged stream reaches of unknown influences across Otago New Zealand. Important findings are twofold. *First*, independent observations of naturalised Median flows compare within few percent of the 50$^{th}$ percentile predictions from the FDC models. *Second*, the naturalised Meta models predict FDCs that outperform the calibrated SWAT model FDCs at gauge sites in the Taieri Freshwater Management Unit: Taieri at Tiroiti, Taieri at Sutton Creek, and Taieri River at Outram. Departures in the naturalised reference state are interpreted as flow regime changes across the duration curves. We believe these Meta models will be useful in predicting naturalised catchment FDCs across other New Zealand regions using physical catchment features available from the national data base.

## 1. Introduction

The flow duration curve (FDC) models the streamflow as a function of exceedance probability or % time flow is equal to or exceeded (Searcy, 1959). Every FDC is dependent upon the temporal interval (e.g., daily, weekly, monthly, or yearly) and catchment scale, e.g., Strahler stream order 1 through 9 (Strahler, 1952), chosen for the analysis. The interval between flow measurements reflects character of the catchment being evaluated, e.g., large rivers (high Strahler order) reflect low gradients with comparatively long-time scales over which flow does not change significantly, or small streams (low Strahler order) reflect high gradients with streamflow measurements that change significantly over comparatively short time scales (Arora et al., 2006). For this reason, the time interval of flow measurements used in computing the FDC is considered with regards to the study needs and data availability.

In general, methods used for computing FDCs reflect applications in gaged and ungauged catchments (Leong and Yokoo, 2021). In *gauged catchments,* the FDC is the complement to a cumulative distribution function derived from flow measurements sorted into class intervals (Vogel and Fennessey, 1994). Other methods developed for gauged settings include nonparametric quantile estimation for quantifying the FDC confidence intervals (Vogel and Fennessey, 1994) and fitting of analytical equations for computing FDCs in semi-arid regions (Ma et al., 2023). While FDCs are commonly constructed for gauged catchments, many hydrologic studies are associated with ungauged catchments. In *ungauged catchments,* researchers rely on developing methods capable of transferring spatial information from gauged to ungauged catchments (Li et al., 2010). For example, FDCs are often transferred from nearby gauges or estimated by a regional FDC using flow normalized by drainage area (discharge) or reference condition (Searcy, 1959). Other statistical techniques for estimating or predicting FDCs include parametric modeling (LeBoutillier and Waylan, 1993; Cigizoglu, 2000; Burgan and Aksoy, 2018, 2020) and nonparametric modeling (Vogel and Fennessy, 1994; Castellarin et al. 2004, 2012). Still others derive FDCs using calibrated equations (Yu et al., 2002; Post, 2004; Lane et al., 2005); and artificial neural network (Atieh et al., 2015, 2017), quasi-Newton (Yaşar and Baykan, 2013), ensemble machine learning (Booker and Woods, 2014), quantile solidarity (Poncelet 2017), and geostatistical (Goodarzi and Vazirian, 2023) methods.

To this point, the statistically and empirically based FDC methods are linear and represent perennial streams. Recently, Burgan and Aksoy (2022) demonstrated the usefulness of the multiple regression technique for estimating FDCs in ungauged subbasins characterized by intermittent streams. In comparing different FDC methods, Ries and Friesz (2000) and Archfield et al. (2007) found empirical methods superior to statistical methods, particularly at mid to high exceedance probabilities (mid to low flows). In another study, Ali and Hasan (2022) found empirical methods superior to a physically based hydrologic model. Despite the number and type of approaches available, few studies compute FDCs for natural streams (Terrier et al., 2021). A natural (or naturalised) stream is one whose ecological description (Acreman and Dunbar, 2014) is defined as pristine condition or minor modification of in-stream and riparian habitat. At regional and national scales, many catchments are likely to be influenced by combinations of human influences (Montanari et al., 2013) that are classified as slightly modified (e.g, individual disturbances associated with water supply schemes or irrigation development), moderately modified (e.g., multiple disturbances associated with dams and diversions), largely modified (e.g., multiple disturbances including dams, diversions, storage, and transfers), seriously modified (high human population density and extensive water resource abstractions).

The *aim* of this study is to develop and test a new machine learning algorithm for predicting naturalized FDCs at human influenced sites across multiple catchment scales. We hypothesize that the combination of natural catchment hydrology and available physical and climate catchment characteristics can provide suitable information for Meta model building and prediction of naturalized FDCs at human influenced catchments. The objectives of this machine learning study are twofold. *First*, a *set* of Meta models will be developed and used to predict naturalized FDCs and their uncertainty at selected exceedance probabilities (one per model) across a range of spatial catchment scales (Strahler order 1 to 7) that include human influenced gauged catchments (N=317) and ungauged river reaches of natural or human influenced conditions (N=18612) across Otago New Zealand. *Second*, the predicted naturalised FDCs are extracted at selected gauge sites and compared to independent observations and results determined using a calibrated SWAT model across the Taieri Freshwater Management Unit (FMU). Departures of naturalised predictions from their human-influenced reference state (Vogel et al., 2007) are used to quantify catchment safe yield (Archfield et al., 2007) and water resource availability (Snelder et al., 2011). This study builds on (1) the work by Friedel et al. (2023) who devised a *single* stacked ensemble machine learning model to predict naturalized hydrology (mean and 7-day mean annual low flow) across gauged catchments and ungauged stream reaches in FMUs of the Otago region; and (2) the work by Rajanayaka et al. (2023) who developed a calibrated SWAT model to simulate naturalised FDCs at selected gauge sites in the Taieri FMU.

## 2. Data and Methods

The proposed algorithm involves four key elements: Data, Base Models, Meta Model, and Final Predictions (Fig. 1). This algorithm is repeated to predict spatial flows at discrete exceedance probabilities that define catchment FDCs from 0 to 100%.

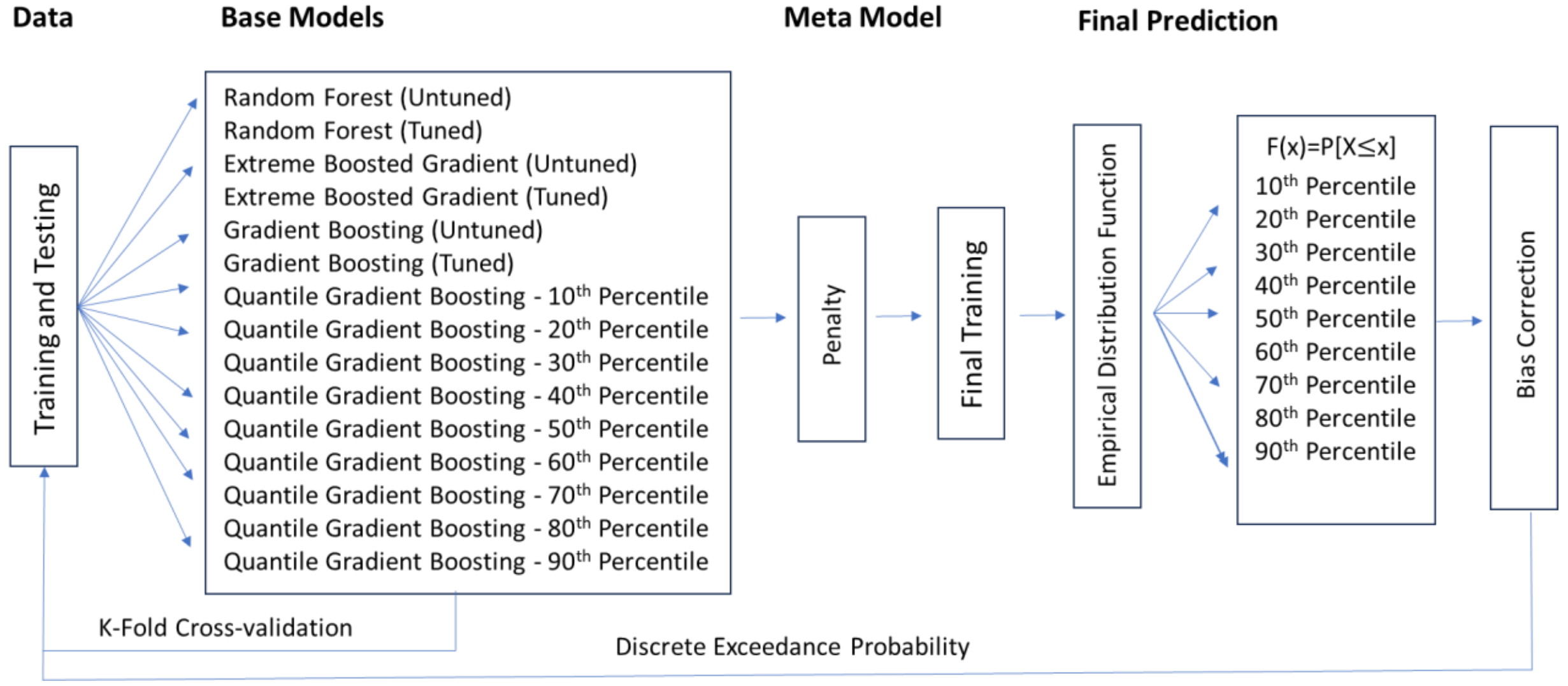

**Fig 1.** Flowchart illustrating the machine learning algorithm used to predict probable naturalised flow duration curves across gauged catchments and ungauged reaches in the Otago Region, New Zealand. This algorithm is repeated to predict flows at discrete exceedance probabilities (separate models) defining the flow duration curve from 0 to 100%.

*2.1 Data*

According to Rallo et al. (2002), one of the elements necessary for accurate training and testing of base models is information diversity. Information diversity reflects the incorporation of training information (response and predictor variables, also called target and features) characterizing mutually informative relations across different spatial and temporal sampling gradients. The types of regression data required include natural flows (*response variables)* at probable exceedances and catchment characteristics *(predictor variables*). The natural streamflow at discrete exceedance values, collectively describing the discrete flow duration curve (FDC), are computed from available daily streamflow time-series (Table 1).

**Table 1**. Flow duration curve derived from observed daily flows, where FDC = flow duration curve and N = the number of discrete exceedance points defining the FDC.

| Index | Description | Calculation |
|---|---|---|
| FDC | Exceedance probability distribution of daily flow | Interpolation of the exceedance proability distribution for daily flows onto N discrete points (from 0 to 100) |

*2.2 Base Models*

The base models used in developing Meta models rely on various ensemble machine learning algorithms (Pedregosa et al., 2011), namely Random Forest Regressor (RFR; Breiman, 2001), Gradient Boosting Regressor (GBR; De'ath, 2007), Extreme Gradient Boosting Regressor (XGB; Chen and Guestrin, 2016), and Quantile Gradient Boosting Regressor (QGBR; Zheng 2012), to predict FDCs. These ensemble algorithms learn relationships among response (streamflow at a selected exceedance probability) and predictor variables (catchment characteristics) without relying on statistical assumptions about the data (Dietterich, 2000). The mathematics associated with these algorithms are briefly reviewed by Friedel et al. (2023). For a detailed review of these methods the reader is referred to the accompanying references. Important base model tasks involve (standard practice) training and testing of the ensemble machine learning models (Dietterich, 2000).

Several decisions are invoked during the base model *training* phase. First, a file with the naturalised catchment records is assigned. Second, the decision is made to assign a discrete exceedance probability as the response variable. Third, the number and type of catchment and climate characteristics are assigned as independent predictor variables. Fourth, an arbitrary random seed (also referred to as the random state) is assigned to initialize the random number generator for shuffling and sampling of the catchment records. Fifth, a decision is made on the relative proportion of records assigned to the training and testing phases. Sixth, a decision is made to assign default ensemble model parameters and/or invoke a hyperparameter tuning method to optimize the model parameter values (Pedregosa et al., 2011).

In ensemble machine learning, hyperparameters refer to input parameters that influence the model structure and their predictions. The available parameters for tuning depend on the type of base model and can determine how closely a model fits the training data. Fitting too closely tends to promote model learning from noise in the training dataset (overfitting). This situation typically results in poor prediction on the testing dataset. Conversely, fitting too loosely means that the model has not learned to represent patterns in the training data (underfitting). Even though there are many approaches to hyperparameter tuning, this study uses simple grid search and random grid search. In using the grid search, the number of values are defined for each parameter, creating a multi-dimensional grid space that includes every combination of hyperparameter values. Consequently, if there is a high number of hyperparameters that require tuning, this approach can become time and computationally expensive. In a random search, the hyperparameter values are sampled from a pre-defined range of values. In both cases, each candidate model is formed on a unique set of hyperparameters, and the best model is chosen as the one that achieves the lowest mean square error on the test dataset.

Each ensemble machine learning algorithm employs a different number of model parameters that may be tuned. For instance, the *Random Forest* method optimizes three parameters during the tuning phase (Scornet 2017). First, the number of variables parameter is randomly selected at each node and considered for splitting. Reducing this parameter increases the randomness of the tree-building process thereby creating trees that are more dissimilar to each other. Second, the number of trees parameter is used to build the forest. Model accuracy typically levels out after arriving at the number of trees required to build a credible model. Third, the tree depth is the point at which the tree stops growing. The larger the tree depth, the closer the model fits the training data increasing the risk of overfitting. In contrast to the Random Forest, the *Gradient Boosting* algorithm uses nine hyperparameters to facilitate convergence to an optimal solution (Malohlava and Candel, 2017). This method implements randomness in the modelling process to avoid overfitting. In addition to number of trees, maximum tree depth, and number of variables sampled for splitting, the number of variables sampled for each tree is also defined by the user. The number of variables sampled at each node is then calculated as the product of the variables sampled for the tree, multiplied by the variables sampled for splitting. The learning rate in this method is the factor by which the contribution of each consecutive tree is reduced compared to the previous tree. Another parameter defines the type of histogram used to speed up selection of the best splitting point at each node. The subsample rate determines the size of the random sample used at each iteration. Smaller samples give rise to lower testing errors whereas larger samples tend to improve the training accuracy. Lastly, there are two hyperparameters that determine the need for additional tree splitting: the minimum required relative improvement in squared error, and the minimum number of observations in a leaf node. Lastly, the *Extreme Gradient Boosting* represents another implementation of the boosting algorithm (Chen and Guestrin, 2016). The number of iterations, the subsample size, maximum tree depth, and fraction of explanatory variables sampled at each tree are also required hyperparameters. In addition, the shrinkage rate determines the learning rate of the algorithm in the training step, i.e. the amount by which the contribution of each consecutive tree is reduced compared to the previous tree. Additional parameters that need tuning when using this algorithm determine how conservative the algorithm is in terms of further partitioning at a leaf node.

The base model *testing* phase is undertaken by presenting the independent split fraction to the trained models. This phase is important for assessing the ability of these models to generalize when presented with independent catchment records. The relative quality of trained regression models is often based on the R-Squared coefficient of determination (Lewis-Beck, 2015) as follows: 60-70% poor, 70-80% good, 80-90% very good, >90% excellent. Scatterplots of predicted values to observed values are often inspected to visually identify prediction bias, where values with a 1:1 correspondence reveals an (ideal) unbiased model. Feature importance scores are sometimes reviewed to evaluate the relative influence that a feature may have on the model prediction process. However, the interpretation of these scores can be misleading because highly correlated features result in splitting their importance giving the false impression that they have less importance. Lastly, deviance plots are inspected to ensure the model is not overfitting the set of training records. Once the training and testing phases are satisfactorily completed, the next step is to create a Meta model.

### *2.3 Meta Model*

A Meta model is created with the aim of achieving greater predictive accuracy by stacking results from multiple base learners (Wolpert, 1992). There are various ways of stacking the low-level base learners into a high-level meta model, e.g., simple weighted average or neural network model to learn the best combination based on the residual errors (Ting and Witten, 2011). In the former case, the weighted average does not consider the quality of models or provide a means for quantifying their predictive uncertainty. In the latter case, there are often issues achieving improvements relating to the learning problem: being well represented by the training data, complex enough that there is more to learn by combining predictions, and choice of base learners are sufficiently uncorrelated in their predictions (or errors). For these reasons, this study implements a new Meta model algorithm to predict flows and their uncertainty. Advantages in using this approach are to prevent overfitting by providing a more robust estimate of the model performance on unseen data and compare different models and select those that perform the best. Disadvantages in using this approach are potentially threefold: an increase in computational time for training when considering multiple folds, time consuming (cross-validation when multiple models need to be compared), and bias-variance tradeoff (choice of the number of randomly shuffled split sets: too few folds may result in high variance, while too many folds may result in high bias).

### *2.4 Final predictions*

The final predictions are made by presenting independent physical catchment features to the retained Meta models. The predictions at gauged and ungauged sites represent a set of Meta models at discrete exceedance probability values where flow is equal to or exceeded (FDC) at values 0 to 100%. At each exceedance probability value, there exists a set of flows at predefined percentiles that are computed as an empirical distribution function (Shorack and Wellner, 1986; Taboga, M., 2021) from the retained model predictions at each site. Extending these measurements across the discrete flow exceedance values provides a complete set of conditional FDCs that quantifies prediction uncertainty at each site. The ability to estimate naturalized probable FDCs provides a basis for discerning likely departures from reference states (Vogel et al., 2007), safe yield (Archfield et al., 2007) and water-resource availability (Snelder et al., 2011) in regional catchments. Comparisons of naturalised probable duration curves to human influenced flow duration curves can be used to quantify the probable rates of decline in flows from their reference state.

## 3. Results and Discussion

### *3.1 Data*

#### 3.1.1 Study Region

The data used in this study are sourced from the Otago Region of New Zealand. In this 32000 km$^2$ region, catchments and river reaches are described as 1$^{st}$ to 7$^{th}$ order streams with areas that range from 0.3 km$^2$ to 6000 km$^2$ (Fig. 2). The Otago Region includes human influenced gauged catchments and ungauged river reaches of which part reflect natural conditions while others reflect human influences of varying intensity. These catchments and river reaches span five Freshwater Management Units (FMUs): Catlins, Clutha (Mata-Au), North Otago, Taieri, Dunedin & Coast. The Clutha (Mata-Au). These FMUs are further subdivided into five smaller water-management units called Rohe reflecting the specialized water-interests of different iwi tribes: Dunstan, Lower Clutha, Manuherekia, and Upper Lakes (Fig. 3).

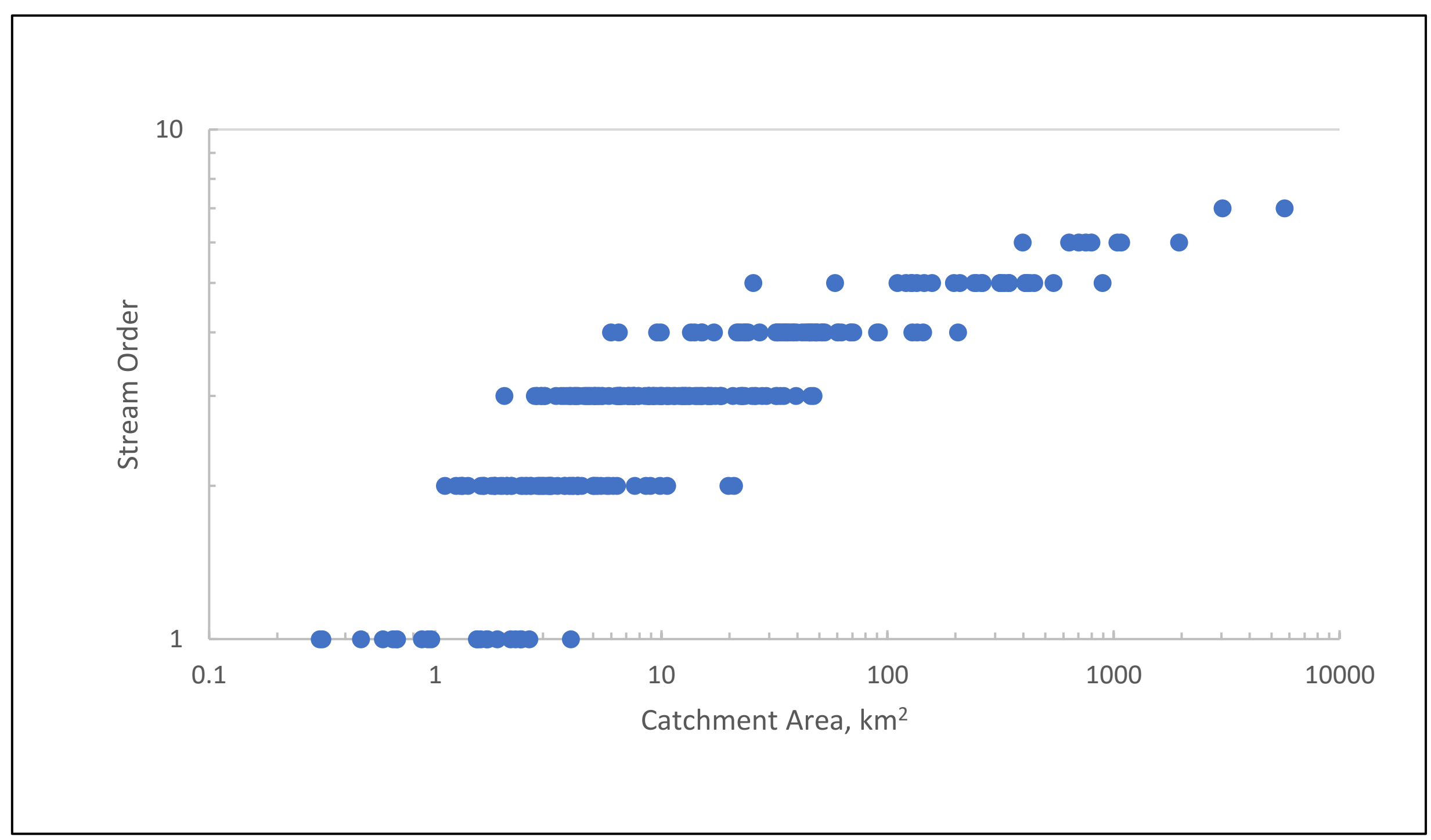

**Fig. 2**. Plot showing the distribution of streamflow gauging sites (blue dots) with respect to the Strahler stream order as function of catchment area in the Otago Region, New Zealand (after Friedel et al., 2023).

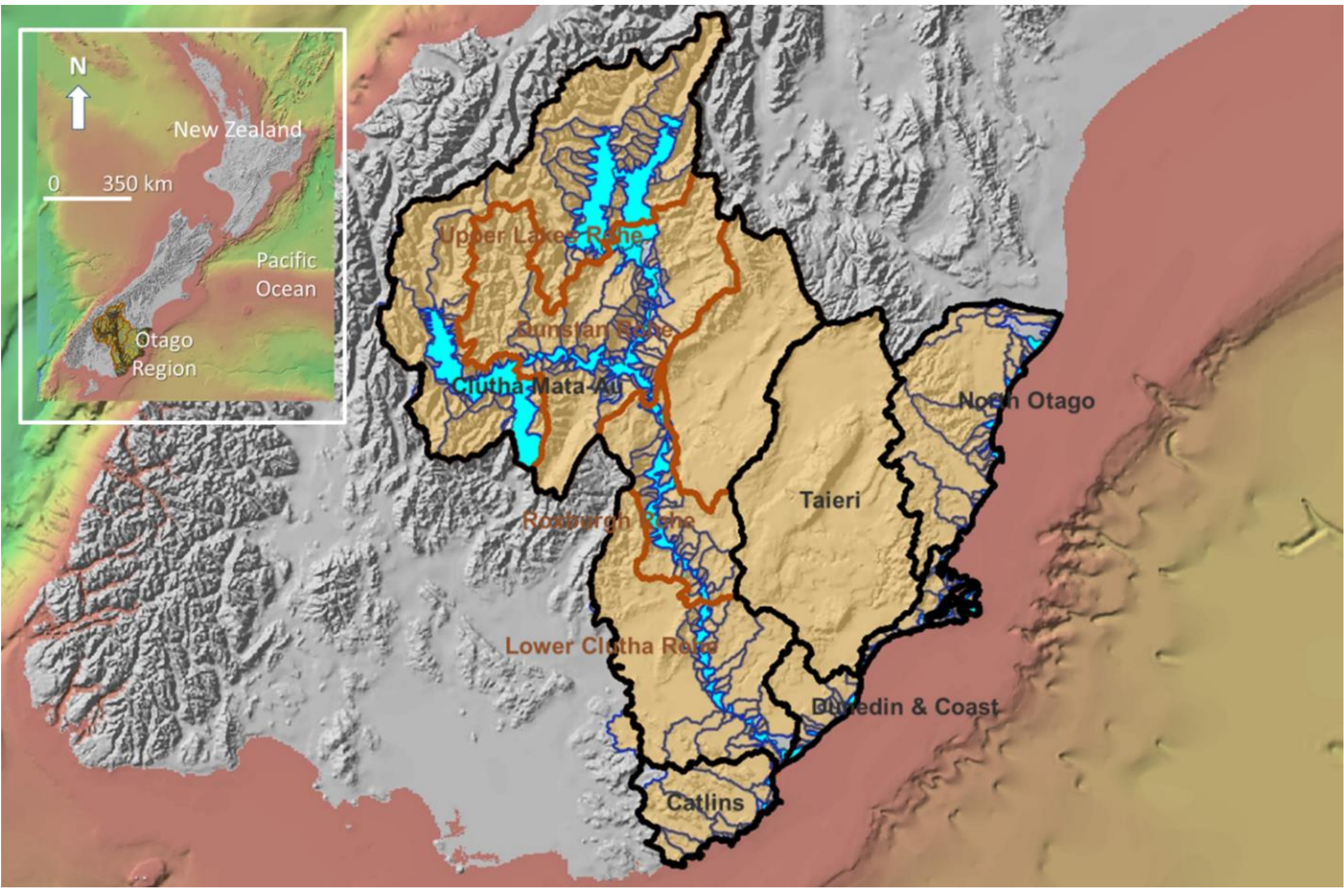

**Fig. 3.** Location map showing water management regions across the Otago Region, New Zealand (after Friedel et al., 2023). The region has five Freshwater Management Units (outlined and labeled in black) that include the Clutha (Mata-Au), Catlins, Dunedin & Coast, North Otago and Taieri. The Clutha comprises five smaller indigenous (iwi) management units (outlined and labeled in brown) called Rohe that include the Dunstan, Lower Clutha, Manuherekia, Roxburgh, and Upper Lakes.

### 3.1.2 Natural Streamflow

The natural streamflow at discrete exceedance values, collectively describing the flow duration curve (FDC), are computed from available daily streamflow time-series collated using the Hilltop software (2023, Hilltop) and Otago Regional Council (ORC) hydrology database. From this database, a set of daily streamflow time-series are collected from gauging stations representing a range of hydrological conditions (natural and human influenced) across the Otago region. Of these sites, only those sites with at least five years of continuous (> 11 months per year) daily flow records are identified for possible use in model building. Additional filtering of time-series records is undertaken to remove gauge stations affected by upstream engineering projects, such as dams, diversions, or substantial abstractions. This last step identified 49 randomly distributed natural streamflow sites for use in model building (Fig. 4). The reader is referred to Booker and Woods (2014) and Friedel et al. (2023) for more details on gauging station selection.

### 3.1.3 Catchment characteristics

There are eight catchment characteristics (features) considered suitable for explaining variation in hydrological patterns across New Zealand (Booker and Snelder, 2012). These eight catchment characteristics include area, elevation, particle size, potential evapotranspiration (PET), rainfall variation, rain days, and runoff volume (Table 2). These characteristics represent median values obtained from the Freshwater Environments of New Zealand geodatabase (Leathwick et al. 2011) sorted on reach numbers found in the River Environment Classification (Snelder and Biggs 2002). The catchment characteristics used in this study represent physical properties located upstream from gauged catchments and ungauged river reaches of mixed environmental conditions. For example, regional catchment characteristics acquired from the locations of 49 natural stream flow sites are presented in Fig. 4. The regional application of trained models use catchment characteristics acquired upstream from 317 human-influenced (named) gauged streamflow sites (Fig. 5), and upstream from 18612 ungauged river reach sites (unnamed) of mixed influence (Fig. 6).

**Table 2.** Summary of physical and climate catchment features (N=8) explaining variation in streamflow exceedance across New Zealand (Leathwick et al. 2011).

| Feature | Description |
| --- | --- |
| Area | Log of catchment area (m2) |
| Elevation | Average elevation in the upstream catchment (m) |
| Particle size | Catchment average of particle size (mm) |
| Potential evapotranspiration (PET) | Annual potential evapotranspiration of catchment (mm) |
| Rainfall variation | Annual catchment rainfall coefficient of variation (mm) |
| Rain days | Catchment rain days, greater than 10 mm/month (days/year) |
| Runoff volume | Percentage annual runoff volume from catchment area with slope > 30° (%) |
| Slope | Average catchment slope (%) |

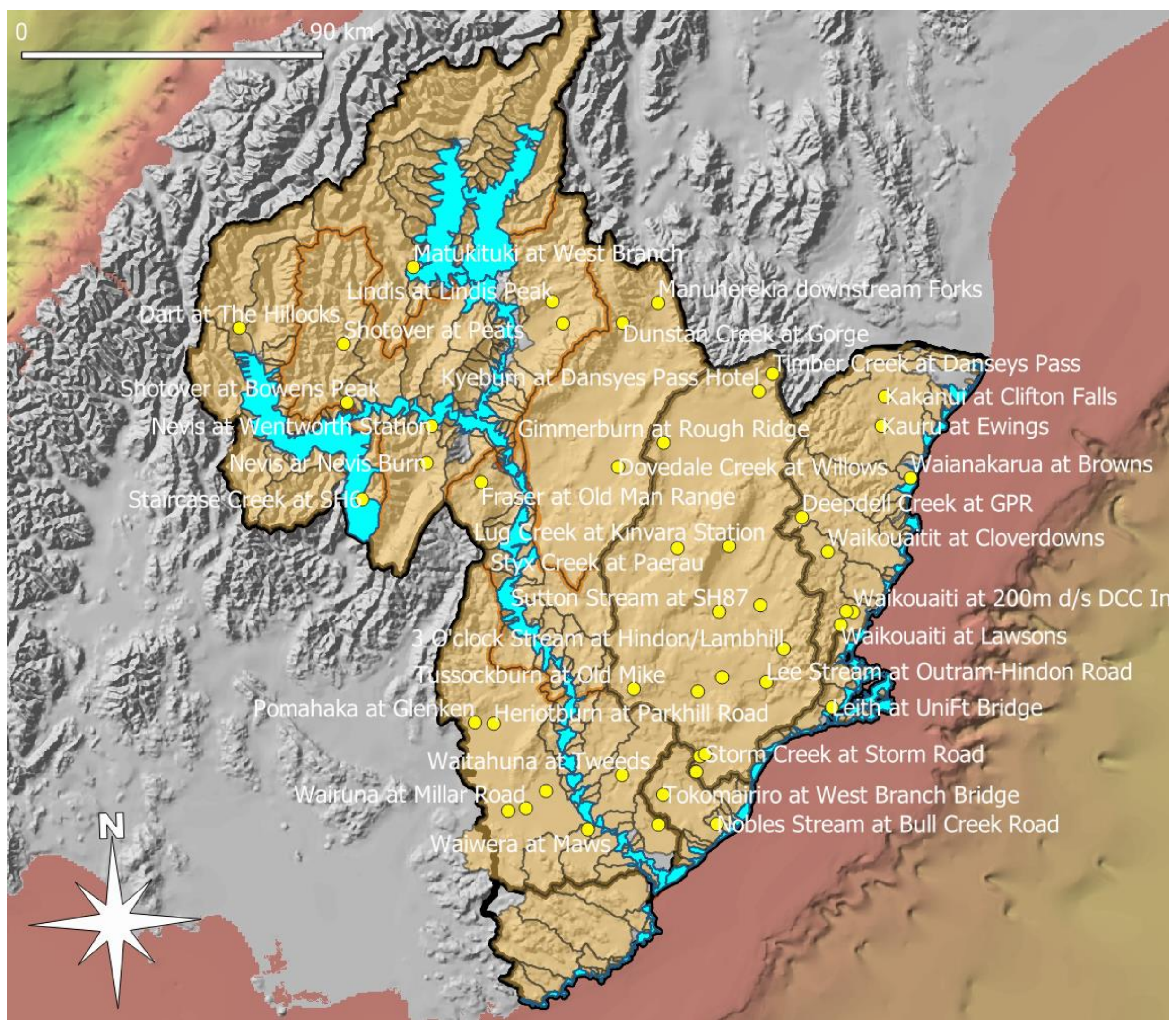

**Fig. 4.** Location map showing names (white text) of randomly distributed gauging stations (yellow dots) that recorded natural flows (N=49) across Otago, New Zealand (after Friedel et al., 2023).

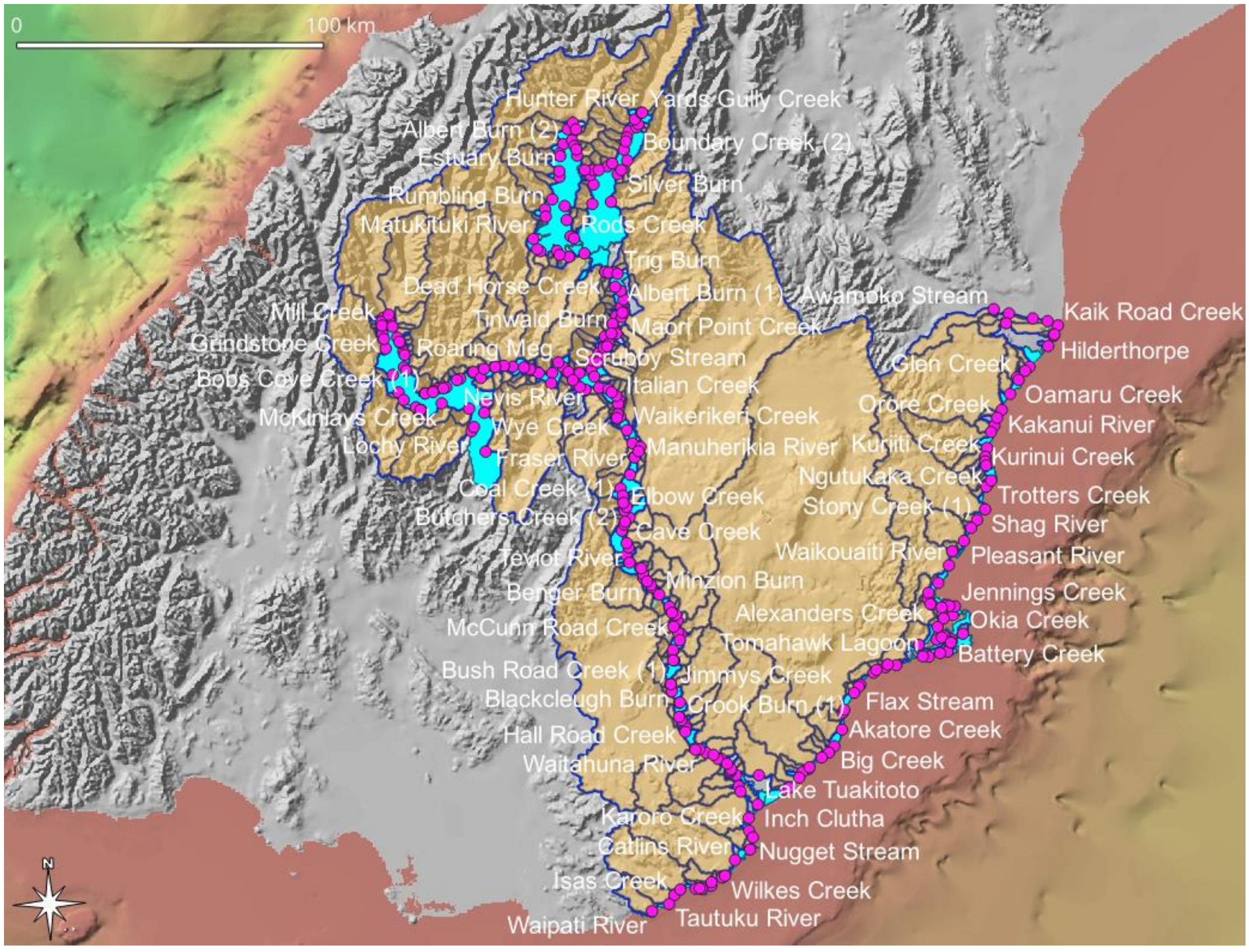

**Fig. 5.** Location map of gauge sites influenced by human activities where the naturalised flow duration curves are predicted at selected (N=317) exceedance percentiles across the Otago Region, New Zealand (after Friedel et al., 2023). The black outlines are the catchment boundaries and purple dots are the streamflow gauge stations influenced by human activities.

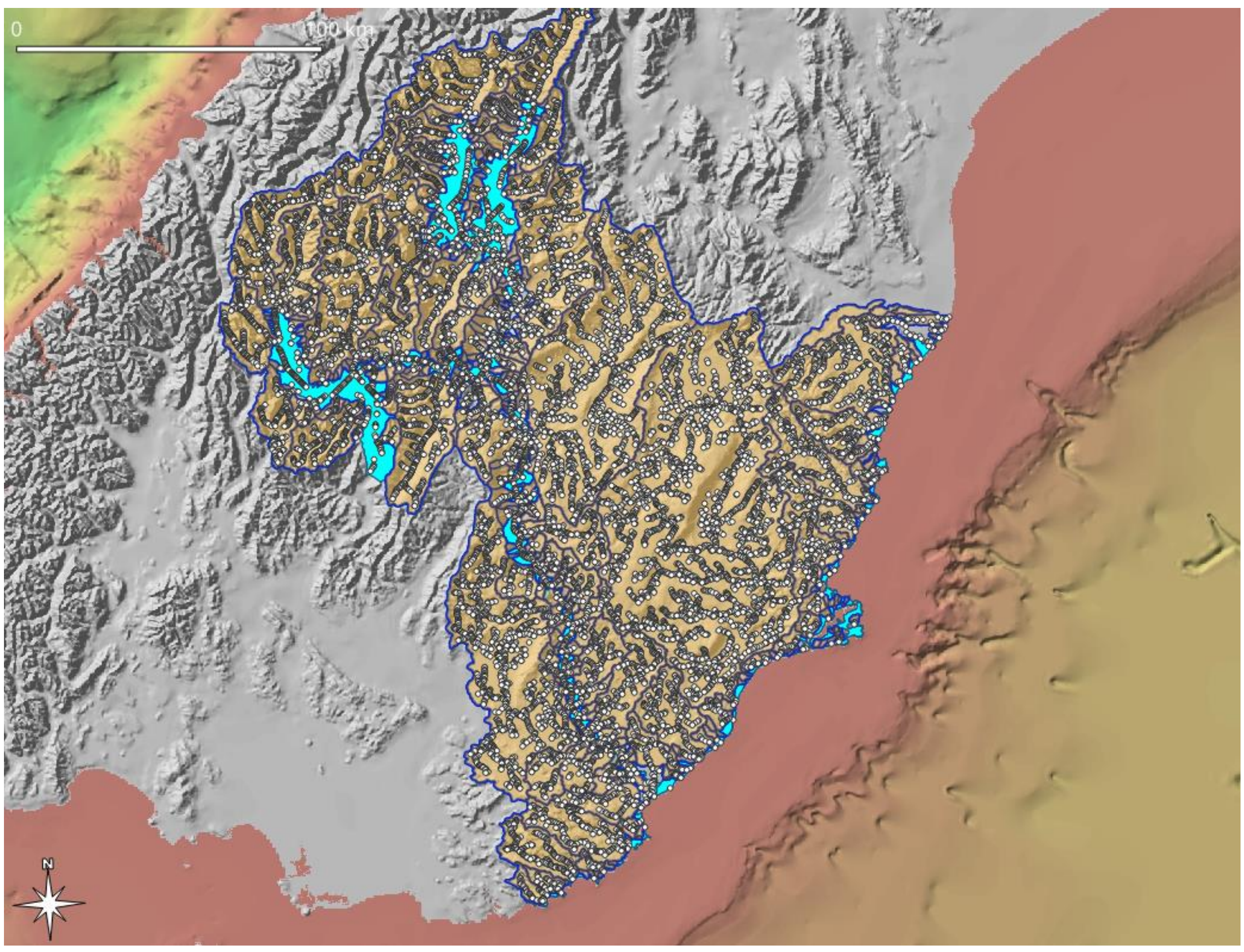

**Fig. 6.** Location map of ungauged river reaches (N=18612) of mixed conditions (natural to human influenced) where the naturalised flow duration curves are predicted at selected percentiles across the Otago Region, New Zealand (after Friedel et al., 2023).

*3.2 Base Models*

In this section, results are provided for model training and testing phases while using catchment records acquired at natural streamflow sites across Otago. Records for a particular catchment site comprise a base model set of flow exceedance probability pairs (response features) and catchment characteristics (predictive features). For each catchment site, the corresponding natural flow duration curve is discretized into twenty five exceedance probability points (e.g., 0, 1, 5, 7, 10, 15, 20, 25, 30, 35, 40, 45, 50, 55, 60, 65, 70, 75, 80, 85, 90, 95, 97, 99, 100%), each of which has eight predictive physical features (see Table 3). These catchment records are randomly shuffled and split five times (K-folds) during the testing and training phase (Dietterich, 2000). In this study, the ratio used in shuffling and splitting the natural flow records is 80% (N=39) for training and 20% (N=10) for testing. A statistical summary of forty-nine natural flow site records that include (independent) catchment characteristics and three (dependent) percent of time discrete exceedance probability values where flow is equal to or exceeded (i.e., (e.g., 0, 50, 100%) is presented in Table 3.

**Table 3.** Summary of catchment characteristic and (selected) flow exceedances (%) models from records at natural streamflow sites used in the base model training and testing phase across the Otago Region.

PET = potential evapotranspiration (mm/unit time), Particle size = mm, FDC0 = Meta model for exceedance probability of 0 (zero percent time flow equals or is exceeded), FDC50 = Meta model of for exceedance probability of 50 (fifty percent time flow equals or is exceeded), FDC100 = Meta model for exceedance probability of 100 (one hundred percent time discharge equals or is exceeded), 50 = fifty percent of time the flow equals or exceeds value (exceedance probability = 50), 100 = one hundred percent of time the flow equals or exceeds value (exceedance probability = 100).

| Meta model: | | | | | | | | | FDC0 | FDC50 | FDC100 |
|---|---|---|---|---|---|---|---|---|---|---|---|
| Exceedance probability (percent of time specified discharges were equaled or exceeded): | | | | | | | | | 0 (%) | 50 (%) | 100 (%) |
| Statistic | Log Area ($m^2$) | Elevation (m) | Partilce Size (mm) | PET (mm/unit time) | Rainfall Variaton (mm) | Rain Days (days/yr) | Runoff Volume (%) | Slope (%) | Flow (m3/s) | Flow (m3/s) | Flow (m3/s) |
| count | 49 | 49 | 49 | 49 | 49 | 49 | 49 | 49 | 49 | 49 | 49 |
| mean | 7.19 | 372 | 3.23 | 935 | 169 | 1.74 | 0.05 | 13.1 | 305 | 2.15 | 0.15 |
| std | 0.85 | 325 | 0.90 | 125 | 16 | 0.67 | 0.10 | 5.64 | 7.17 | 6.45 | 8.62 |
| min | 5.87 | 14.8 | 1.10 | 404 | 143 | 0.81 | 0.00 | 0.33 | 3.00 | 0.07 | 0.002 |
| 25% | 6.57 | 126 | 2.57 | 885 | 155 | 1.44 | 0.01 | 9.66 | 86.0 | 0.45 | 0.02 |
| 50% | 7.05 | 214 | 3.51 | 958 | 168 | 1.65 | 0.02 | 13.3 | 250 | 1.61 | 0.12 |
| 75% | 7.53 | 569 | 3.90 | 995 | 179 | 1.87 | 0.04 | 15.3 | 916 | 11.2 | 0.83 |
| max | 9.50 | 1180 | 4.79 | 1166 | 203 | 5.35 | 0.48 | 28.4 | 37374 | 54.8 | 6.1 |

Selecting a subset of catchment records for base model training and testing is controlled by assigning a random state (also called the random number or random seed) which initiates record shuffling prior to splitting. This process is repeated to produce different subsets (one per random state) of target hydrologic indices and feature catchment characteristics that are presented to the suite of base models. In this way, the shuffling process provides a means to evaluate the effect of different catchment subsets on the prediction uncertainty (bias plus variance) of the base models despite their limited number of records. Another benefit in using this approach is that each random state produces a single reproducible (deterministic) outcome that can be repeated using the same python script for review and/or use in other related analyses at any time. In this study, different randomly shuffled split sets are used to train each of the base models along with variants reflecting the application with and without hyperparameter tuning available from the scikit-learn python toolkit (Pedregosa et al., 2011).

Hyperparameter tuning includes random grid search and random grid search plus cross-validation (e.g., cross-validated Hypertuning involves fitting 5 folds for each of 90 candidates, totaling 450 fits) methods available from this toolbox. In total there are 5 random states evaluated for each of 16 base models giving 80 total base models evaluated during the training and testing phase. This process is extended then to each of the 24 discrete flow exceedance values (e.g., 0, 1, 3, 5, … 90, 95, 97, 99, 100 %) comprising the FDC resulting in 1920 base models that are available to define each gauged and ungauged site FDC. Given the enormous amount of training and testing output, the supporting tables and figures are restricted to those used in predicting probable naturalised flows at the 50$^{th}$ exceedance probability (FDC50). For example, a summary of base model testing quality for this point on the FDC is indicated by R-squared ($R^2$) values for each of the 5 random states in Table 4. Selected scatterplots associated with random state 5 and FDC50 are presented along with their $R^2$ values in Fig. 7. Of the 80 potential base models, only 55 preferred minimum variance models ($R^2 \geq 0.7$) are retrained by presenting the full suite of catchment records.

**Table 4.** Summary of base model quality for the 50% exceedance probability (FDC50) point as indicated by their respective R-squared values for each of the 5 random states, where RFR = Random Forest Regressor, RFRgs = Random Forest Regressor with grid search Hypertuning, XGR = Extreme Gradient Boosting Regressor, XGRgscv = Extreme Gradient Boosting Regressor with grid search and cross-validation Hypertuning, GBR = Gradient Boosting Regressor, GBRgs = Gradient Boosting Regressor with grid search Hypertuning, QGBR10 = Qantile Gradient Boosting Regressor at the 10th percentile, QGBR20 = Qantile Gradient Boosting Regressor at the 20th percentile, QGBR30 = Qantile Gradient Boosting Regressor at the 30th percentile, QGBR40 = Qantile Gradient Boosting Regressor at the 40th percentile, QGBR50 = Qantile Gradient Boosting Regressor at the 50th percentile, QGBR60 = Qantile Gradient Boosting Regressor at the 60th percentile, QGBR70 = Qantile Gradient Boosting Regressor at the 70th percentile, QGBR80 = Qantile Gradient Boosting Regressor at the 80th percentile, and QGBR90 = Qantile Gradient Boosting Regressor at the 90th percentile.

| FDC50 Number | Base Model | Random State 1 | 2 | 3 | 4 | 5 | Available |
|---|---|---|---|---|---|---|---|
| 1 | RFR | 0.85 | 0.55 | 0.84 | 0.80 | 0.71 | 5 |
| 2 | RFRgs | 0.91 | 0.69 | 0.92 | 0.83 | 0.87 | 5 |
| 3 | RFRgscv | 0.81 | 0.58 | 0.78 | 0.74 | 0.84 | 5 |
| 4 | XGB | 0.89 | 0.54 | 0.92 | 0.77 | 0.66 | 5 |
| 5 | XGBgscv | 0.91 | 0.60 | 0.95 | 0.80 | 0.64 | 5 |
| 6 | GBR | 0.91 | 0.65 | 0.89 | 0.75 | 0.71 | 5 |
| 7 | GBRRgs | 0.88 | 0.74 | 0.73 | 0.71 | 0.71 | 5 |
| 8 | QGBR10gs | -0.01 | 0.73 | 0.27 | -0.64 | -0.26 | 5 |
| 9 | QGBR20gs | 0.25 | 0.55 | 0.61 | 0.59 | 0.74 | 5 |
| 10 | QGBR30gs | 0.76 | 0.73 | 0.86 | 0.83 | 0.81 | 5 |
| 11 | QGBR40gs | 0.78 | 0.75 | 0.90 | 0.81 | 0.87 | 5 |
| 12 | QGBR50gs | 0.72 | 0.67 | 0.89 | 0.89 | 0.88 | 5 |
| 13 | QGBR60gs | 0.78 | 0.66 | 0.75 | 0.82 | 0.75 | 5 |
| 14 | QGBR70gs | 0.87 | 0.65 | 0.70 | 0.80 | 0.82 | 5 |
| 15 | QGBR80gs | 0.86 | 0.59 | 0.39 | 0.73 | 0.63 | 5 |
| 16 | QGBR90gs | 0.60 | -0.97 | -1.79 | 0.50 | -0.32 | 5 |
| | | | | | | Total = | 80 |

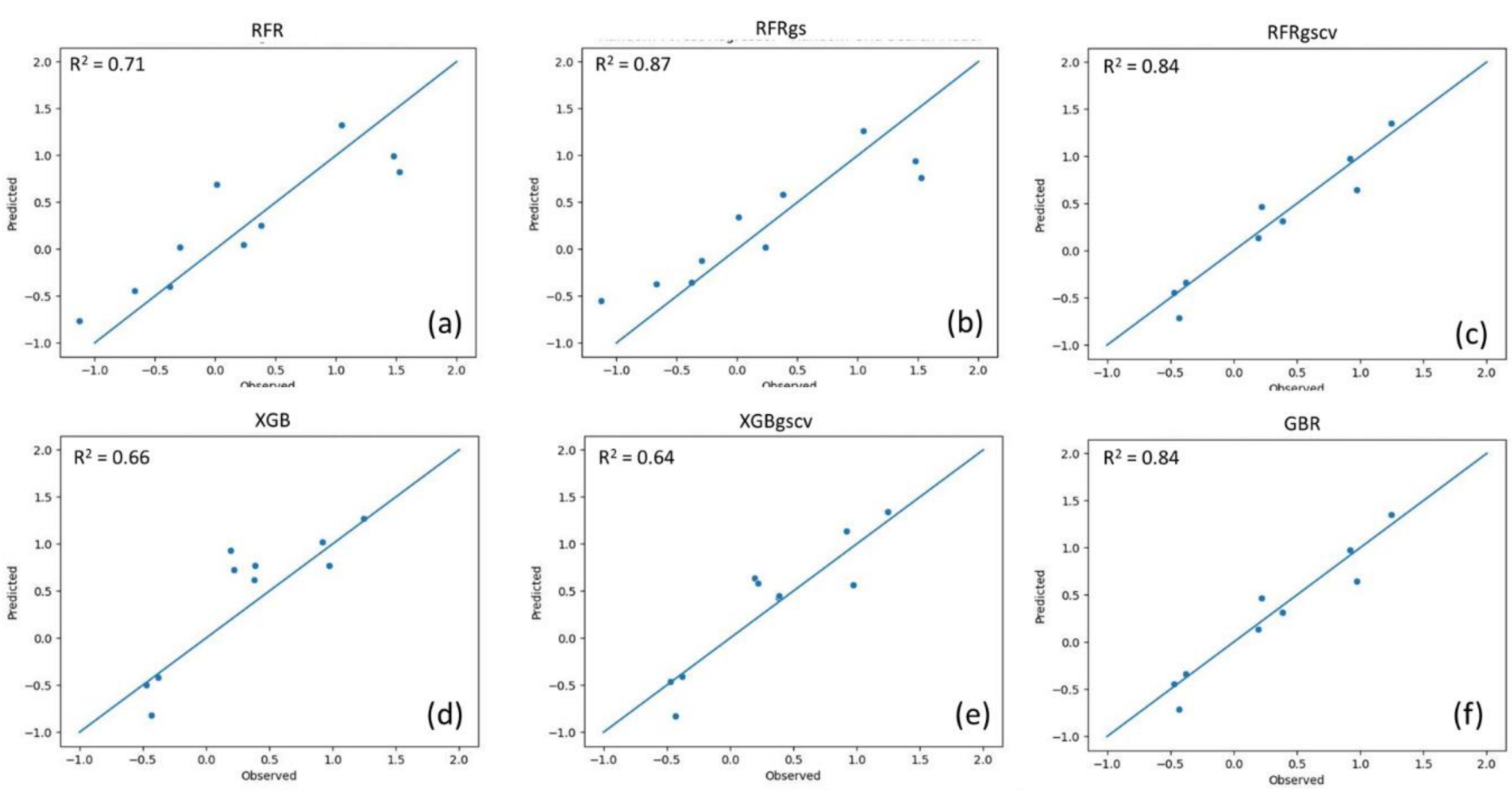

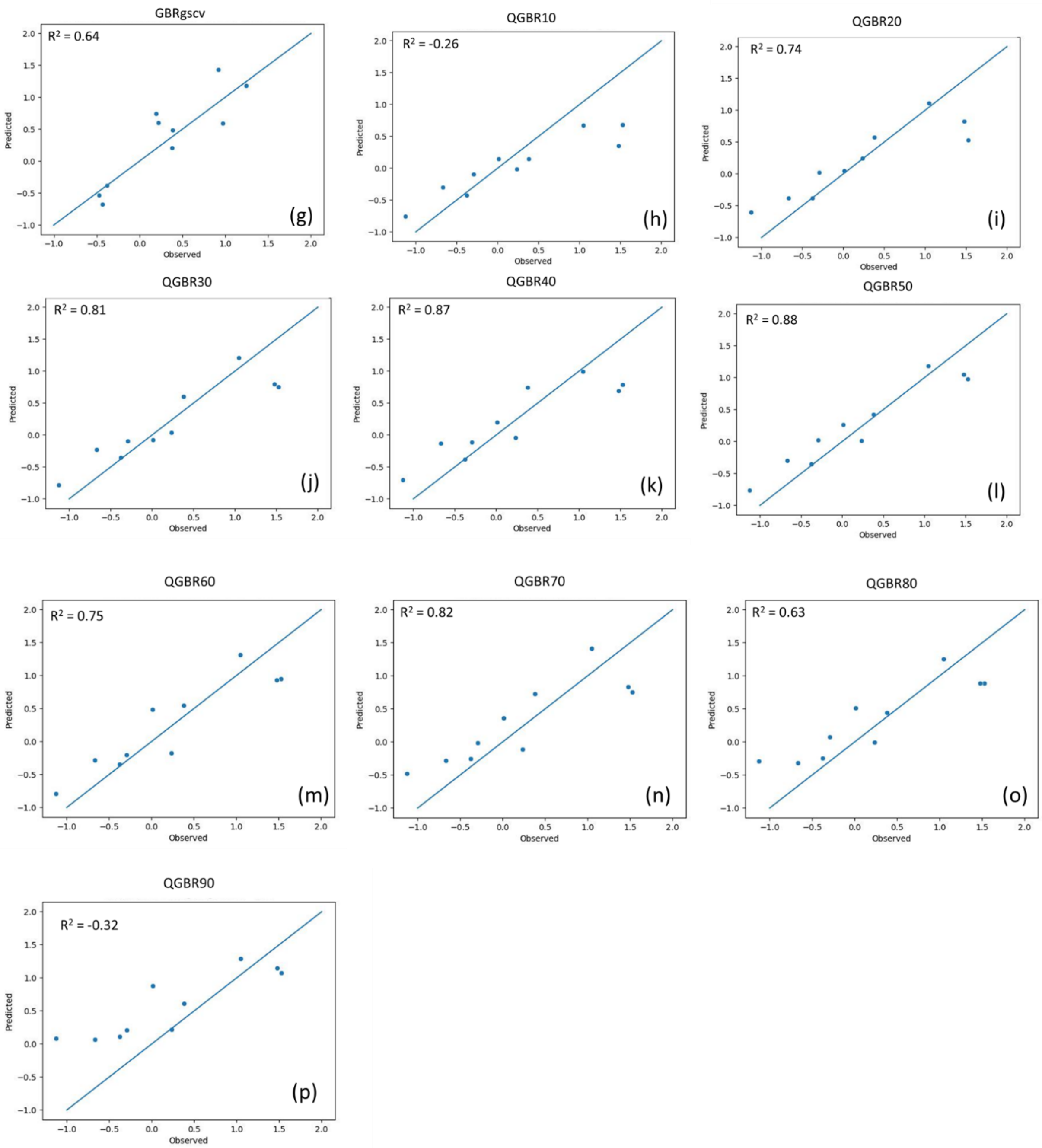

**Fig. 7.** Scatterplots for base model testing phase (N=16) using different random states (N=5) at the 50% exceedance probability (FDC50): (a) RFR, (b), RFRgs, (c) RFRgscv, (d) XGB, (e) XGBgscv, (f) , GBR (g) GBRgscv, (h) QGBR10, (i) QGBR20 , (j) QGBR30, (k) QGBR40, (l) QGBR50, (m) QGBR60, (n) QGBR70, (o) QGBR80, (p) QGBR90. RFR = Random Forest Regressor, RFRgs = Random Forest Regressor with grid search Hypertuning, XGR = Extreme Gradient Boosting Regressor, XGRgscv = Extreme Gradient Boosting Regressor with grid search and cross-validation Hypertuning, GBR = Gradient Boosting Regressor, GBRgs = Gradient Boosting Regressor with grid search Hypertuning, QGBR10 = Qantile Gradient Boosting Regressor at the 10th percentile, QGBR20 = Qantile Gradient Boosting Regressor at the 20th percentile, QGBR30 = Qantile Gradient Boosting Regressor at the 30th percentile, QGBR40 = Qantile Gradient Boosting Regressor at the 40th percentile, QGBR50 = Qantile Gradient Boosting Regressor at the 50th percentile, QGBR60 = Qantile

Gradient Boosting Regressor at the 60th percentile, QGBR70 = Qantile Gradient Boosting Regressor at the 70th percentile, QGBR80 = Qantile Gradient Boosting Regressor at the 80th percentile, and QGBR90 = Qantile Gradient Boosting Regressor at the 90th percentile.

### *3.3 Meta Models*

The Meta models are created with the aim of achieving greater accuracy through reduced predictive uncertainty by stacking results from the base models (learners). In this study, the algorithm used to develop Meta models relies on four steps. *First*, a discrete flow exceedance probability value is chosen, e.g. 0, 1, 3, 5, 10, 15, … 85, 90, 95, 97, 99, or 100%. *Second*, the K-fold cross-validated base models with satisfactory metrics are retained, e.g., penalty: $R^2 \geq 0.7$ (good, very good, and excellent models). For example, implementing this penalty for the 50% exceedance probability point (FDC50) reduces the pool of stacking candidates from 80 base models to 55 base models. *Third*, the retained base models are retrained using the available natural flow and physical property records without splitting. *Fourth*, these steps are repeated to produce a set of regional Meta models that correspond to the chosen exceedance probability values across Otago.

This algorithm results in a set of unstacked base models that are used to predict probable naturalised flows at discrete exceedance probabilities associated with site flow duration curves. For example, the probable naturalised flow predictions for the 0%, 50%, and 100 % exceedance probability models are presented as a function of random state and area at 317 human influenced sites in Fig. 8. Inspecting this figure reveals a positive prediction bias (over predicted) for flows in catchments less than about 1 km$^2$ (Strahler stream order of 1), and negative prediction bias (under predicted) for flows in catchments greater than about 1000 km$^2$ (Strahler order 6 and 7). The general trend in these biases appear consistent and independent of the base model chosen.

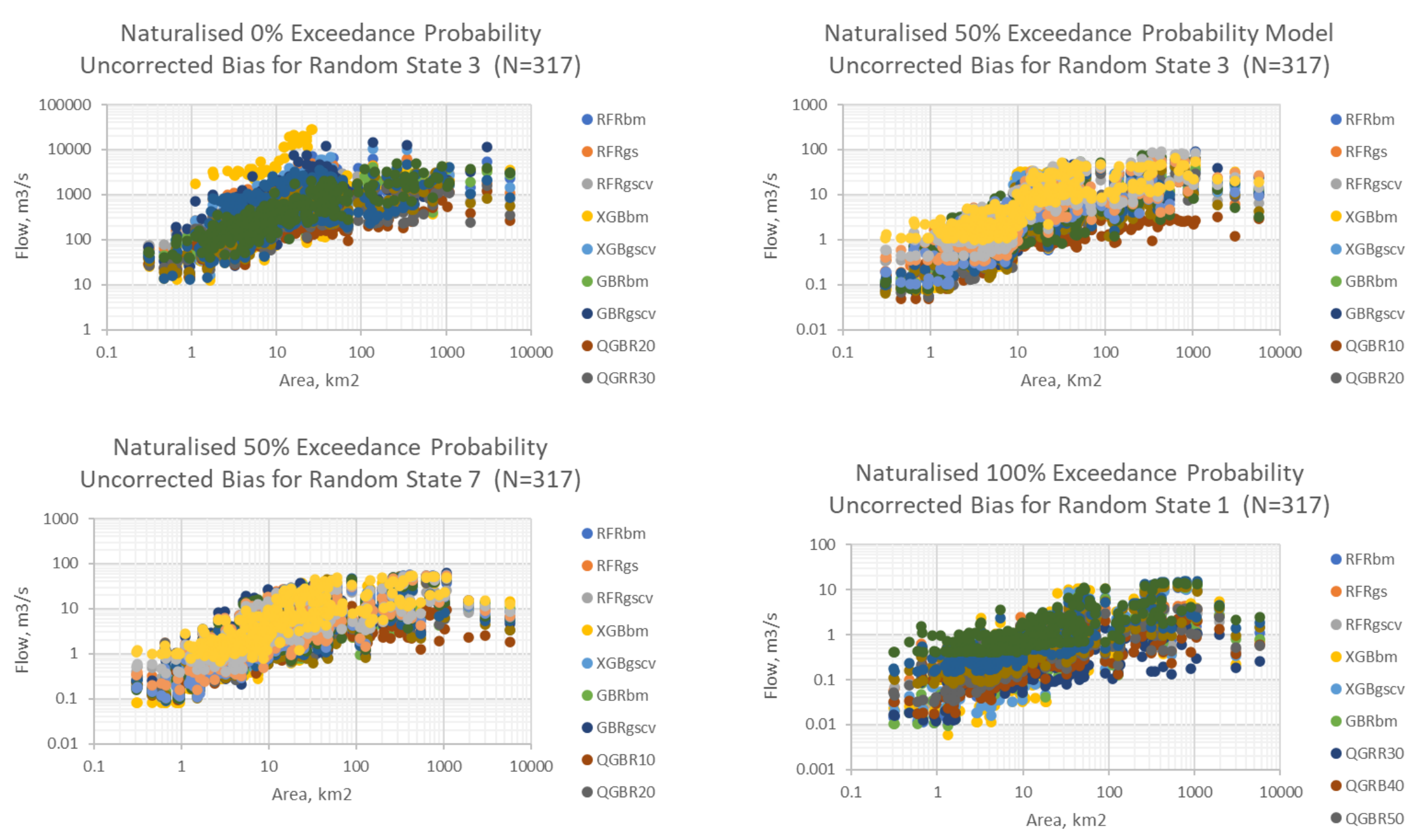

**Fig 8.** Scatterplots reveal biased (base model) naturalised flow predictions as function of gauged catchment area (N=317). (a) Naturalised uncorrected flows for 0% exceedance probability models with random state = 3, (b) Naturalised uncorrected flows for 50% exceedance probability models with random state = 3, (c)

Naturalised uncorrected flows for 50% exceedance probability models with random state = 7. Prediction of naturalised flows appear biased downward (under predicted) for catchment areas greater than 1000 $km^2$ (Stahler order 7 and 8) and biased upward (over predicted) for catchment areas less than 1 $km^2$ (Stahler order 1). RFR = Random Forest Regressor, RFRgs = Random Forest Regressor with grid search Hypertuning, XGR = Extreme Gradient Boosting Regressor, XGRgscv = Extreme Gradient Boosting Regressor with grid search and cross-validation Hypertuning, GBR = Gradient Boosting Regressor, GBRgs = Gradient Boosting Regressor with grid search Hypertuning, QGBR10 = Qantile Gradient Boosting Regressor at the $10^{th}$ percentile, QGBR20 = Qantile Gradient Boosting Regressor at the $20^{th}$ percentile, QGBR30 = Qantile Gradient Boosting Regressor at the $30^{th}$ percentile, QGBR40 = Qantile Gradient Boosting Regressor at the $40^{th}$ percentile, QGBR50 = Qantile Gradient Boosting Regressor at the $50^{th}$ percentile, QGBR60 = Qantile Gradient Boosting Regressor at the $60^{th}$ percentile, QGBR70 = Qantile Gradient Boosting Regressor at the $70^{th}$ percentile, QGBR80 = Qantile Gradient Boosting Regressor at the $80^{th}$ percentile, and QGBR90 = Qantile Gradient Boosting Regressor at the $90^{th}$ percentile.

## 3.4 Final Predictions

The final prediction of naturalised flows and their prediction uncertainty at discrete exceedance probabilities along the FDC requires further model enhancements. These enhancements include correcting prediction bias and validating prediction results.

### *3.4.1 Bias Corrections*

Quantifying the *prediction uncertainty* is accomplished by computing an empirical distribution function from the set of meta model predictions (Shorack and Wellner, 1986). For example, the scatterplots of naturalised flows at various prediction deciles are presented as conditional distributions for the 50% exceedance probability model as a function of catchment area (Fig. 9). The left panel in this figure reveals an upward prediction bias across uncertainty (e.g, $10^{th}$, $20^{th}$, …, $80^{th}$, and $90^{th}$ deciles) for flows in catchments with areas less than about 1 $km^2$, and downward prediction bias across uncertainty for flows in catchments with areas greater than about 1000 $km^2$. These findings are consistent with the prediction bias identified in the individual (unstacked) base models (see Fig 8). Potential reasons for the tendency to predict biased flows may be attributed to: (1) the minimization of prediction variance (as opposed to bias) when using ensemble machine learning as base models, and/or (2) the incorrect number of randomly shuffled split sets where too few folds may result in high variance, while too many folds may result in high bias. The application of a linear *bias correction* across decile model predictions, as opposed to applying corrections to meta models at discrete random states. is shown in the right panel of this figure. This bias correction procedure is applied to each of the remaining stacked Meta models resulting in cumulative probability distribution functions conditioned by area. These distribution functions reveal noise at lower flow percentiles for areas between about 5 to 100 $km^2$.

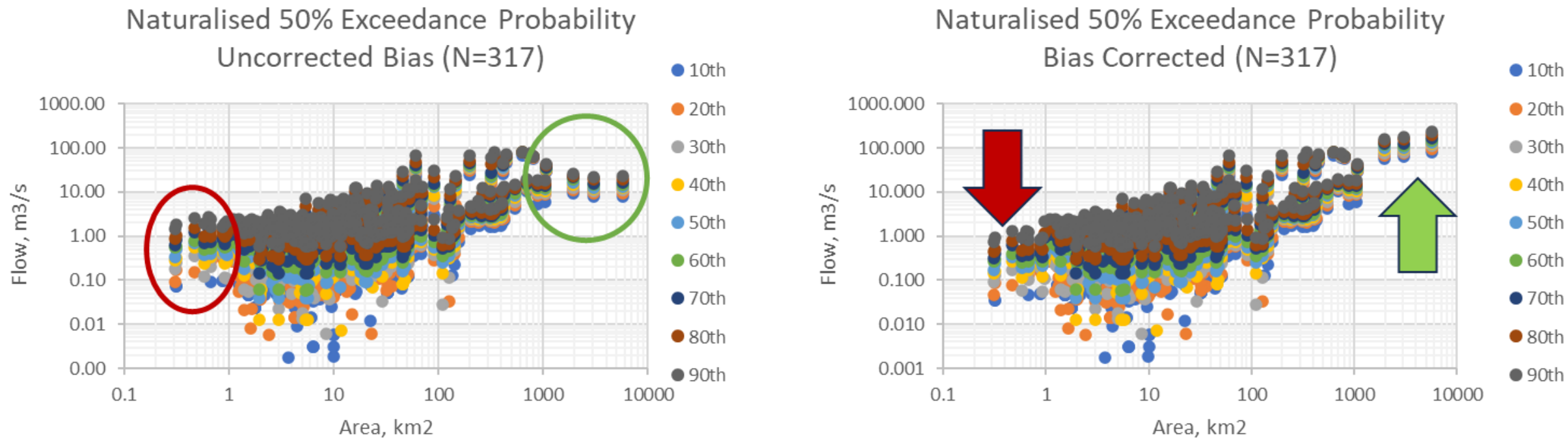

**Fig. 9.** Scatterplot of probable naturalised flows as a function of area for the 50% exceedance probability model. The left panel reveals upward prediction bias for flows in catchments less than about 1 km$^2$, and downward prediction bias for flows in catchments greater than about 1000 km$^2$. The right panel shows the application of simple linear bias correction.

*3.4.2 Validation*

This section provides validation results to assess the relative quality of Meta model predictions. *First* the bias corrected median (50$^{th}$ percentile) natural predictions at the 50% exceedance probability is compared to the natural median flows (observations) computed from gauged measurements recorded at the regional Dart River natural flow site (635.2 km$^2$; Strahler order 6; 23-year period of record from 6/12/1996) and the regional Matukituiki River natural flow site (799.3 km$^2$; Strahler order 6; 41-year period of record from 8/21/1979 through 7/15/2020). For example, consider the histograms of naturalised flow predictions at the Dart and Matukituki River sites (Fig. 10). Streamflow at these river sites are not included in the development of Meta models, rather these flows are predicted by presenting their independent physical catchment properties to the bias corrected Meta model. For example, the respective observed Median flow at the Dart River gauge sites is 80 m3/s and compares closely with the Median prediction value of 80.2 m3/s. Likewise, the observed Median flow at the Matukituki River gauge site is determined to be 62 m3/s and compares closely with the median prediction value of 62.7 m3/s. The very good fit at these two natural flow sites may be attributed to their comparatively large catchment areas (high Strahler order) reflecting low gradients with comparatively long-time scales over which flow does not change significantly.

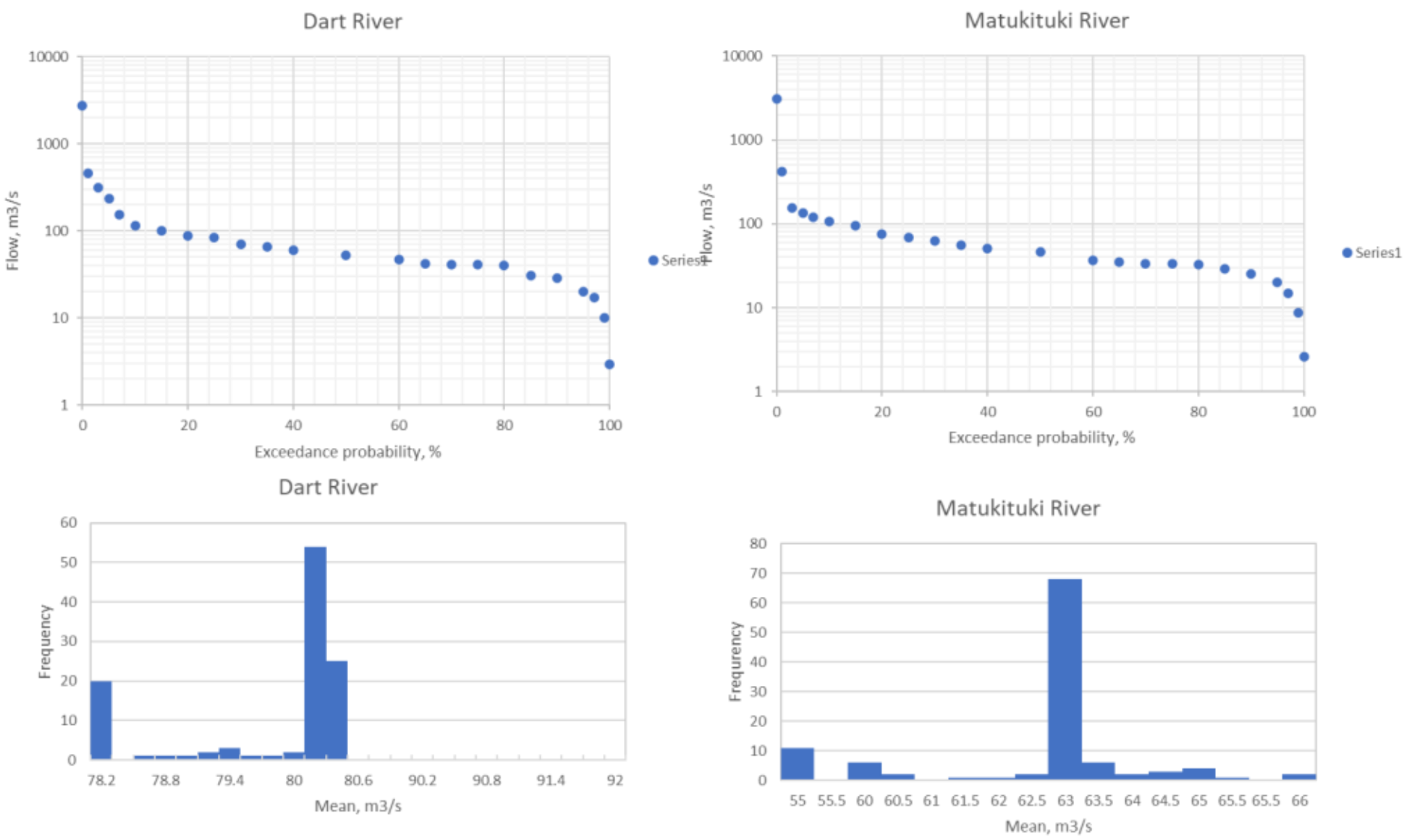

**Fig. 10**. Naturalised flow duration curves (upper panels) and histograms of naturalised flow predictions (lower panels) for Dart ($6^{th}$ order stream covering 648 $km^2$) and Matukituki River ($6^{th}$ order stream covering 816 $km^2$) sites. Streamflow at these sites are not included in the development of Meta models (independent). The respective Median flow predicted as $50^{th}$ percentile flow duration curves at the Dart River and Matukituki River gauge sites are 53.2 m3/s, (median observed value is 54.8 m3/s) and 45.8 m3/s (median observed value is 44.1 m3/s). The respective Mean flow predicted at the Dart River and Matukituki River gauge sites are 80 m3/s (Mean observed value is 80.2 m3/s) and 62 m3/s (Mean observed value is 62.7 m3/s). Measured streamflow at the Dart River is about 23 years over the period from 6/12/1996 through 1/13/2021, whereas streamflow at the Matukituki River is about 41 years over the period from 8/21/1979 through 7/15/2020.

*3.4.3 Application*

The training, testing and validation of Meta models supports their use in predicting FDCs. This section presents the analysis of naturalised Meta model predictions across the Otago Region and the Taieri Freshwater Management Unit. To do so requires only that the independent catchment characteristics be presented to the final Meta models.

*3.4.3.1 Otago Region*

A summary of regional naturalised flow duration predictions at selected exceedance probabilities (percent time flow equaled or exceeded), prediction percentile (uncertainty), and Strahler stream order is presented in Table 6. These results are aggregated by stream order demonstrating the range in naturalised flows among Strahler stream order at 0% exceedance probability (high flows), 50% exceedance probability (median flows), and 100% exceedance probability (low flows) at different prediction deciles. As expected, there is an increase in the range of naturalised flows with increasing Strahler stream order (e.g., from 1 to 7) and increasing prediction decile (e.g., $10^{th}$, $50^{th}$, $90^{th}$). At any exceedance probability, these results demonstrate that there is a range in prediction uncertainty (e.g., difference in prediction at the $90^{th}$ and $10^{th}$ deciles or $75^{th}$ and $25^{th}$ percentiles) at each Strahler Stream Order in the Otago region. For example, at the $7^{th}$ stream order the range in prediction uncertainty at the $0^{th}$ exceedance probability (high flow) reflects a prediction range of 1165-5376 m3/s to 1964-10306 m3/s, and at the $7^{th}$ stream order the range in prediction uncertainty at the 100% exceedance probability (low flow) is 1.34-2.76 to 3.02 to 5.65 m3/s. The standard interpretation holds that at the 90$^{th}$ percentile the range of flows will be equal to or less than the predicted range in flows, at the 50$^{th}$ the percentile the range of flows will be equal to or less than the predicted range in flows, and at the 10$^{th}$ the percentile the range of flows will be equal to or less than the predicted range in flows. In considering the 50$^{th}$ percentile (median) the flows also increase from with increasing Strahler order but decrease with increasing exceedance probability along the flow duration curve (from high flows to low flows). In summary, there is a probable range of regional naturalised flows predicted at each Strahler order that is dependent on prediction decile and discrete location along the FDC.

One interpretation of this table could be that at the current state, the 50$^{th}$ percentile (Median flow) the expected range in flows at the 0% exceedance probability (peak flows) for the 1$^{st}$, 2$^{nd}$, 3$^{rd}$, 4$^{th}$, 5$^{th}$, 6$^{th}$, and 7$^{th}$ order streams will be in the range of 24-177 m3/s, 177-333 m3/s, 333-622 m3/s, 622-1031 m3/s, 1031-1571 m3/s, 1571-1974 m3/s, 1964-100306 m3/s; at the 50% exceedance probability (intermediate flows) for the 1$^{st}$, 2$^{nd}$, 3$^{rd}$, 4$^{th}$, 5$^{th}$, 6$^{th}$, and 7$^{th}$ order streams will be in the range of 0.17-1.0 m3/s, 1.0-1.8 m3/s, 1.8-5.0 m3/s, 5.0-9.0 m3/s, 9.0-14.7 m3/s, 14.7-27.1 m3/s, 27.1-46.6 m3/s; and at the 100% exceedance probability (low flows) for the 1$^{st}$, 2$^{nd}$, 3$^{rd}$, 4$^{th}$, 5$^{th}$, 6$^{th}$, and 7$^{th}$ order streams will be in the range of 0.01-0.17 m3/s, 0.17-0.4 m3/s, 0.4-0.64 m3/s, 0.64-0.86 m3/s, 0.86-1.29 m3/s, 1.29-2.2 m3/s, 2.2-3.91 m3/s; and shift from naturalised to human influenced across duration curves reflects degradation across the flow regime in contrast to partial degradation determined at the larger streamflow sites.

**Table 6**. Summary of naturalised flow predictions at selected exceedance probabilities (percent time flow equaled or exceeded), prediction percentile (uncertainty), and Strahler stream order.

| Exceedance Probability: | 0% | 0% | 0% | 50% | 50% | 50% | 100% | 100% | 100% |
|---|---|---|---|---|---|---|---|---|---|
| Percentile: | 10th | 50th | 90th | 10th | 50th | 90th | 10th | 50th | 90th |
| Strahler Stream Order | High Flow (m3/s) | High Flow (m3/s) | High Flow (m3/s) | Median Flow (m3/s) | Median Flow (m3/s) | Median Flow (m3/s) | Low Flow (m3/s) | Low Flow (m3/s) | Low Flow (m3/s) |
| 7 | 1165-5379 | 1964-10306 | 6866-1003732 | 17.4-40.5 | 27.1-46.6 | 37.3-56.5 | 1.34-2.76 | 2.2-3.91 | 3.02-5.85 |
| 6 | 854-1165 | 1571-1964 | 4177-6866 | 8.5-17.4 | 14.7-27.1 | 23.7-37.3 | 0.78-1.34 | 1.29-2.2 | 2.19-3.02 |
| 5 | 535-853 | 1031-1571 | 2873-4177 | 5.5-8.5 | 9.0-14.7 | 13-37.3 | 0.54-0.78 | 0.86-1.29 | 1.65-2.19 |
| 4 | 340-535 | 622-1031 | 1623-2873 | 3.1-5.5 | 5.1-9.0 | 8.9-13 | 0.38-0.54 | 0.64-0.86 | 1.12-1.65 |
| 3 | 169-340 | 333-622 | 836-1623 | 1.1-3.1 | 1.8-5.0 | 5.1-8.9 | 0.23-0.38 | 0.4-0.64 | 0.7-1.12 |
| 2 | 96-169 | 177-333 | 459-836 | 0.48-1.1 | 1.0-1.8 | 2.4-5.1 | 0.085-0.23 | 0.17-0.4 | 0.34-0.7 |
| 1 | 8-96 | 24-177 | 67-459 | 0.08-0.48 | 0.17-1.0 | 0.3-2.4 | 0.01-0.085 | 0.01-0.17 | 0.01-0.34 |

*3.4.3.2 Taieri Freshwater Management Unit*

In this section, results are presented in the form of flow exceedance probability maps and flow duration curves in (and around) the Taieri FMU. The Taieri FMU is a regional catchment that covers about 5706 km$^2$ (see Fig. 3). These results rely on extracting information from the naturalised stochastic flow duration curves predicted at 18612 reach sites of which 3000 sites are in the Taieri FMU. In this case, the flow values are extracted at Taieri FMU sites by decile and exceedance probabilities comprising the flow duration curves. From these data, selected maps are generated to demonstrate the types of information available to support the Otago Regional Land and Water Plan required under the National policy statement for freshwater management (Ministry for the Environment, 2011, 2015, 2020).

Maps are presented illustrating the spatial distribution of naturalised flows predicted across Strahler stream order in the Taieri FMU at the 0% exceedance probability (percent time flows equaled or exceeded) and the 10$^{th}$, 50$^{th}$, and 90$^{th}$ percentiles (Fig. 11). These three panels capture the range in prediction uncertainty at the high flow exceedance probability across the Taieri FMU and adjacent area. Another set of maps are presented illustrating the spatial distribution of naturalised flows predicted across Strahler stream order in the Taieri FMU at the 0%, 50% and 100% exceedance probability (percent time flows equaled or exceeded) and 50th percentile (Fig. 12). These three panels capture median flows across the peak-flow, mid-flow, and low-flow regimes across the Taieri FMU and adjacent area.

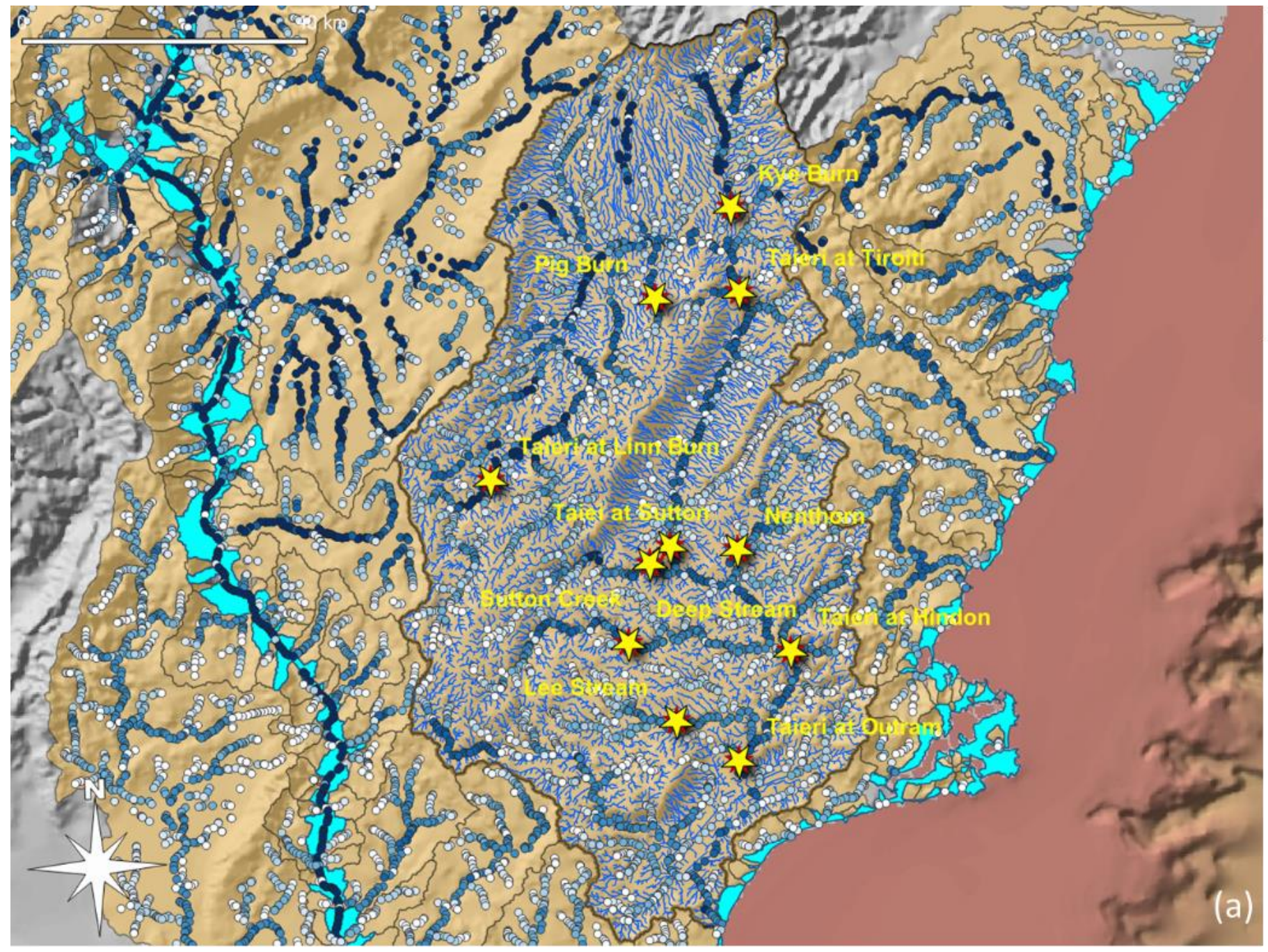

| Exceedance Probability: | | 0% |
|---|---|---|
| Percentile: | | 10th |
| Strahler Stream Order | | High Flow (m3/s) |
| 7 | • | 1165-5379 |
| 6 | • | 854-1165 |
| 5 | • | 535-853 |
| 4 | • | 340-535 |
| 3 | • | 169-340 |
| 2 | ○ | 96-169 |
| 1 | ○ | 8-96 |

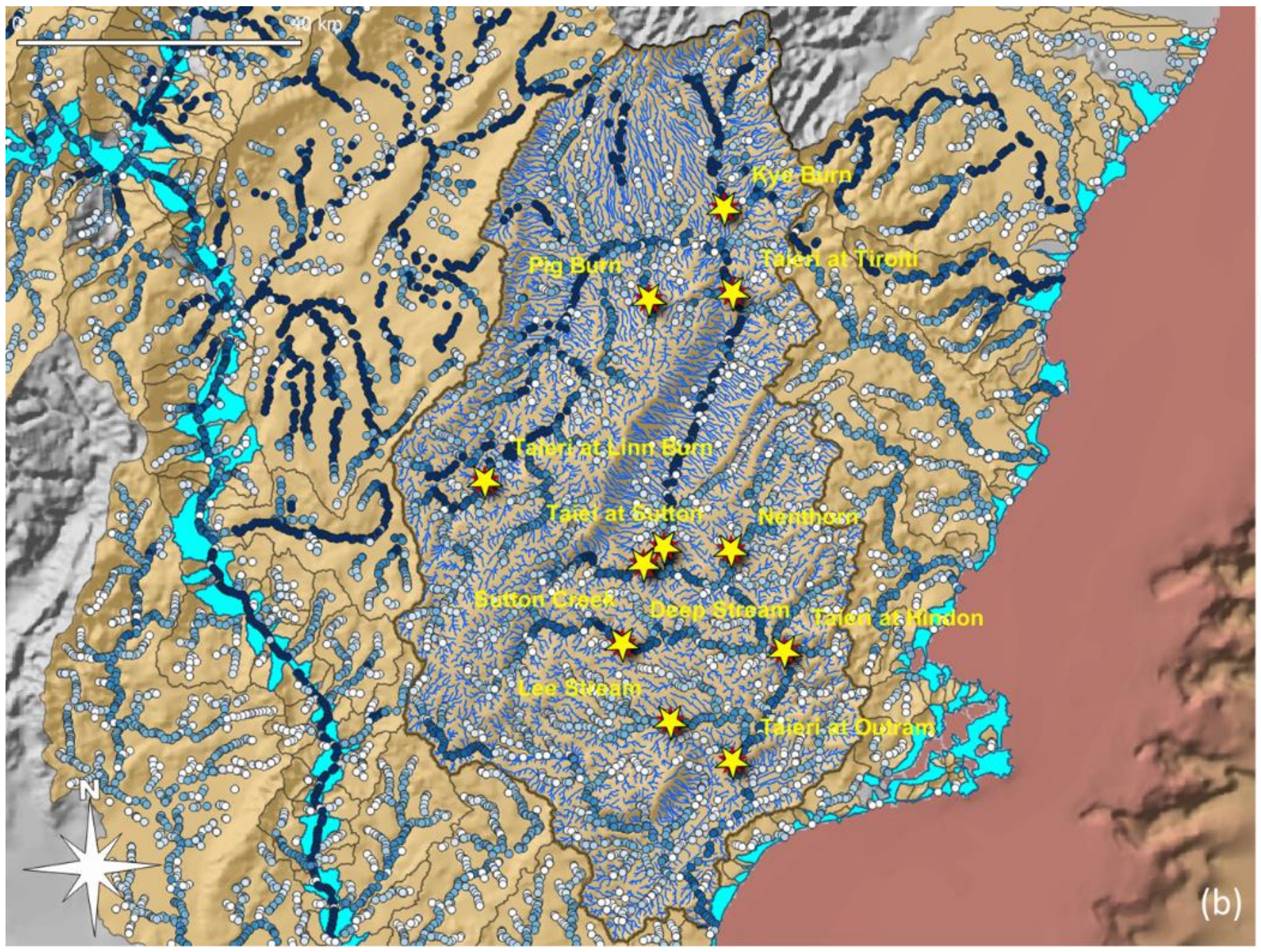

| Exceedance Probability: | | 0% |
|---|---|---|
| Percentile: | | 50th |
| Strahler Stream Order | | High Flow (m3/s) |
| 7 | • | 1964-10306 |
| 6 | • | 1571-1964 |
| 5 | • | 1031-1571 |
| 4 | • | 622-1031 |
| 3 | • | 333-622 |
| 2 | ○ | 177-333 |
| 1 | ○ | 24-177 |

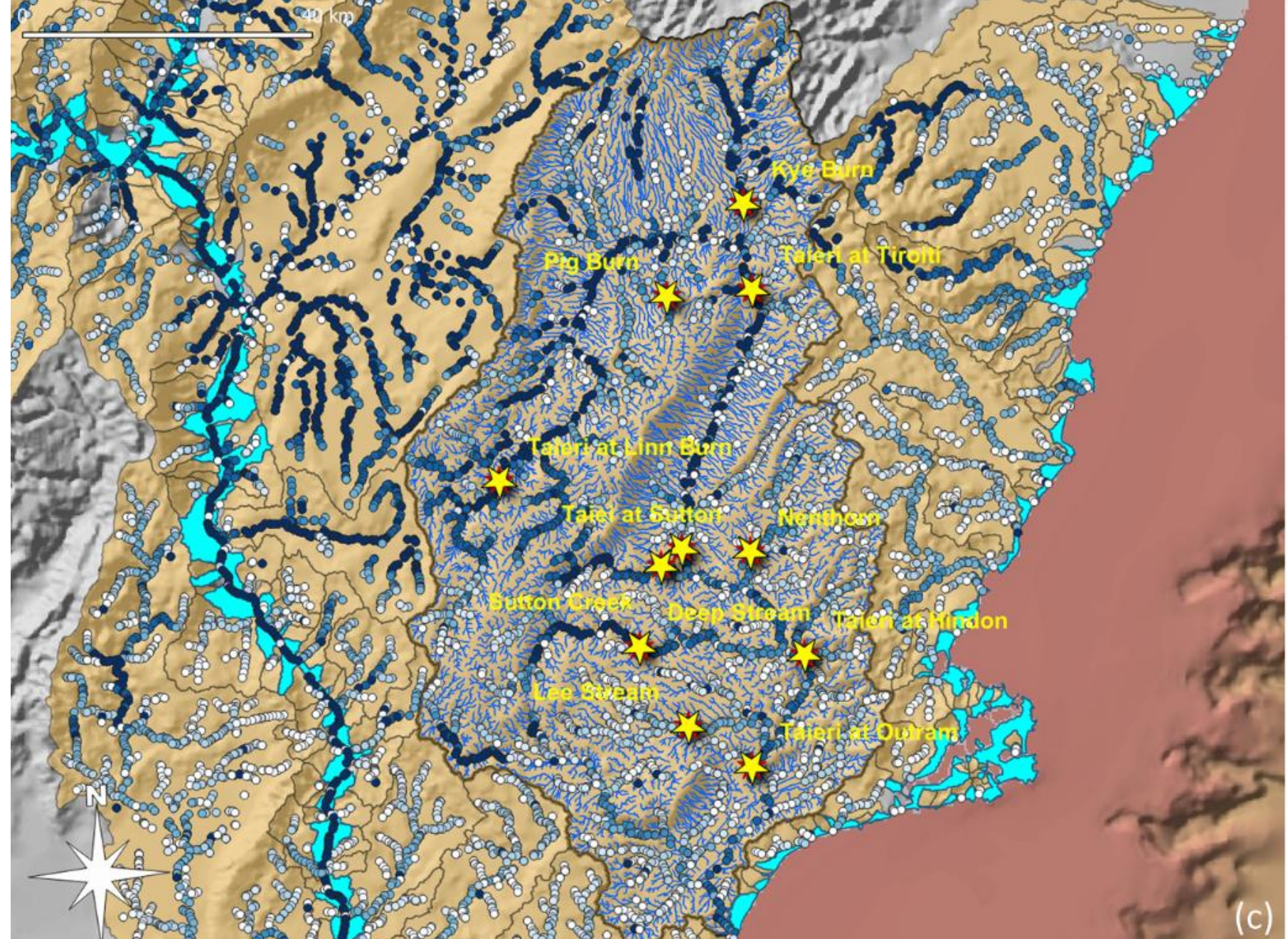

| Exceedance Probability: | 0% |
| --- | --- |
| Percentile: | 90th |
| Strahler Stream Order | High Flow (m3/s) |
| 7 | 6866-1003732 |
| 6 | 4177-6866 |
| 5 | 2873-4177 |
| 4 | 1623-2873 |
| 3 | 836-1623 |
| 2 | 459-836 |
| 1 | 67-459 |

**Fig. 11.** Maps illustrating the spatial distribution of uncertainty in naturalised high flow predictions across Strahler stream order in the Taieri FMU at the 0% exceedance probability (percent time flows equaled or exceeded) and the 10th, 50th, and 90th percentiles. (a) Flows at the 0% exceedance probability and 10th percentile. (b) Flows at the 0% exceedance probability and 50th percentile (median). (c) Flows at the 0% exceedance probability and 90th percentile. These three panels capture the range in prediction uncertainty across the Taieri FMU and adjacent area.

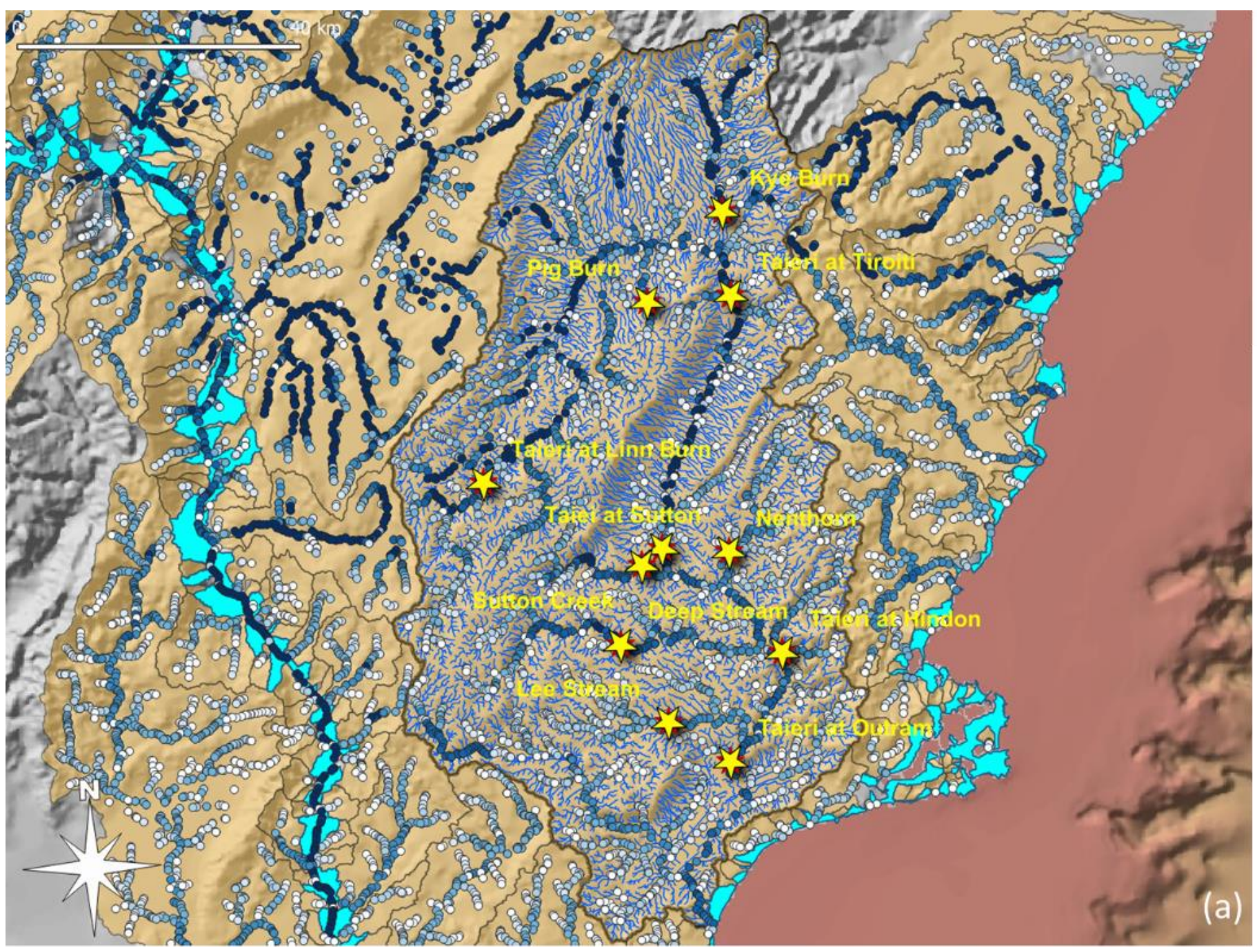

| Exceedance Probability: | 0% |
| --- | --- |
| Percentile: | 50th |
| Strahler Stream Order | High Flow (m3/s) |
| 7 | 1964-10306 |
| 6 | 1571-1964 |
| 5 | 1031-1571 |
| 4 | 622-1031 |
| 3 | 333-622 |
| 2 | 177-333 |
| 1 | 24-177 |

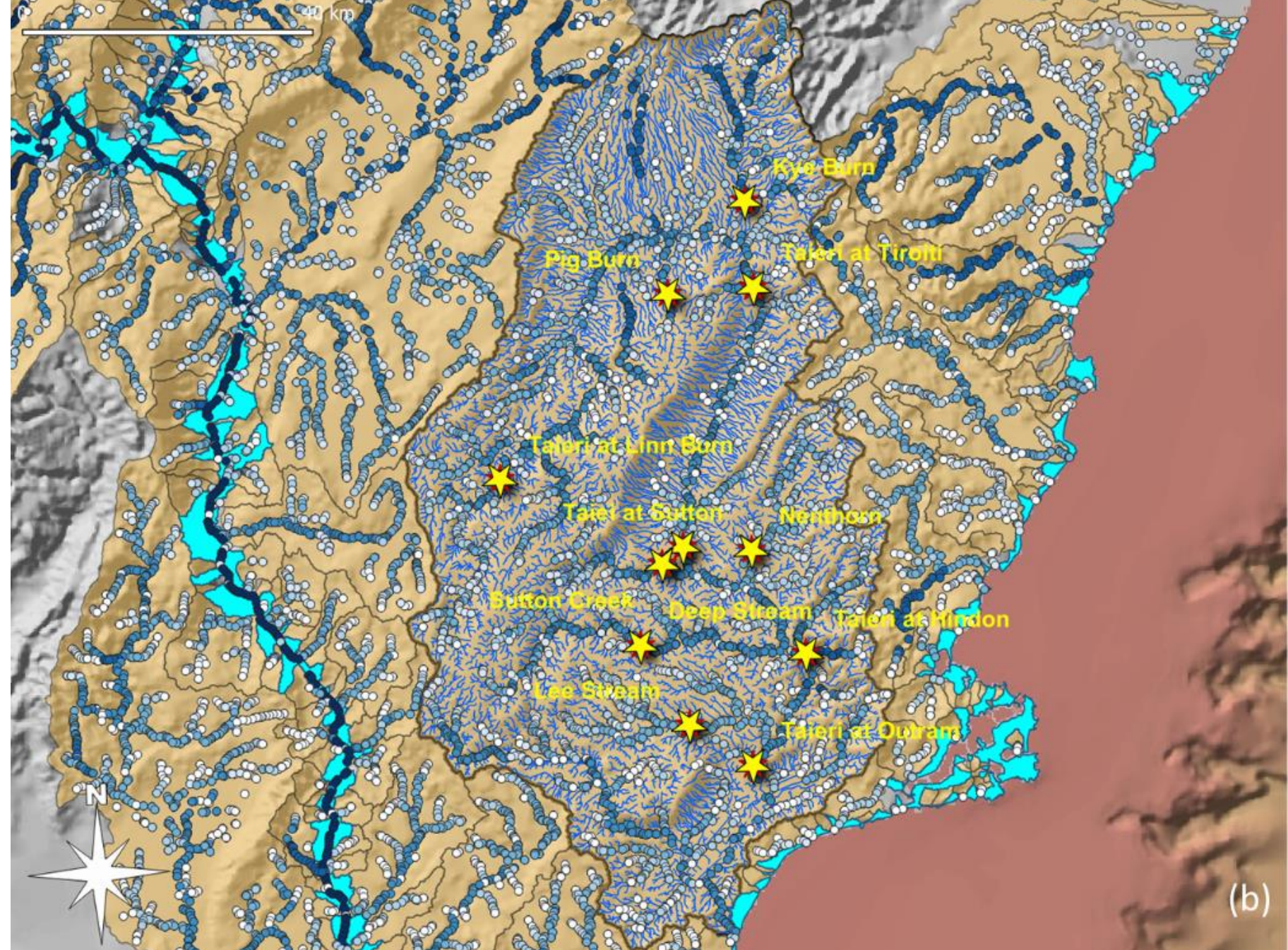

| Exceedance Probability: | | 50% |
|---|---|---|
| Percentile: | | 50th |
| Strahler Stream Order | | Median Flow (m3/s) |
| 7 | • | 27.1-46.6 |
| 6 | • | 14.7-27.1 |
| 5 | • | 9.0-14.7 |
| 4 | • | 5.1-9.0 |
| 3 | ○ | 1.8-5.0 |
| 2 | ○ | 1.0-1.8 |
| 1 | ○ | 0.17-1.0 |

| Exceedance Probability: | | 100% |
|---|---|---|
| Percentile: | | 50th |
| Strahler Stream Order | | Low Flow (m3/s) |
| 7 | • | 2.2-3.91 |
| 6 | • | 1.29-2.2 |
| 5 | • | 0.86-1.29 |
| 4 | • | 0.64-0.86 |
| 3 | ○ | 0.4-0.64 |
| 2 | ○ | 0.17-0.4 |
| 1 | ○ | 0.01-0.17 |

**Fig. 12.** Maps illustrating the spatial distribution of naturalised flows predicted at the 50th percentile (median) across Strahler stream order in the Taieri FMU at three exceedance probabilities (percent time flows equaled or exceeded): 0% (high-flow magnitudes), 50% (middle-flow magnitudes) and 100% (low-flow magnitudes). (a) Flows at the 50th percentile for the 0% exceedance probability point. (b) Flows at the 50th percentile for the 50% exceedance probability point. (c) Flows at the 50th percentile for the 100% exceedance probability point. These three panels capture the median flow predictions at the peak-flow, middle-flow, and low-flow regimes across the Taieri FMU and adjacent area.

In addition to maps, individual FDCs can be plotted for any gauge or ungauged reach site in the region. This is possible by extracting FDCs by prediction percentile from nodes on the stream network closest to the regional station coordinates. If network nodes are not collocated with regional coordinates, these values are considered approximate with FDCs predicted at larger catchments (high Strahler order) more certain reflecting low gradients with comparatively long-time scales over which flow does not change significantly, whereas predictions extracted at smaller streams (low Strahler order) may be less certain reflecting high gradients with streamflow measurements that change significantly over comparatively short time scales. For example, the predicted naturalised FDCs are presented at the 50$^{th}$ percentile for selected gauge sites in the Taieri FMU (Fig. 13). Inspecting this figure reveals naturalised FDC profiles of differing magnitudes.

The naturalised flow regime at these sites was previously unknown and important for understanding their sustainability (Hayes et al., 2021). In general, gauge sites with larger stream catchments have profiles that are associated with larger flows across the discrete exceedance probabilities. A summary of naturalised flow duration predictions (50$^{th}$ percentile) for these Taieri FMU sites at three exceedance probabilities (where 0% = high flow, 50% = median flow, 100% = low flow) is summarized in Table 7. Inspecting values at the 50 percent exceedance probability reveals a continuum of flows with the largest associated with the Taieri at Outram site and the smallest associated with the Nenthorn at Mt Stoker Road site. Inspecting flows across FDCs reveal subtle changes associated with streams of similar catchment areas. For example, the Lee and Nenthorn sites are similar in character across their FDCs but depart at lower flows (exceedance probabilities greater than about 80%) indicating the former stream may be more resilient to droughts than the later stream.

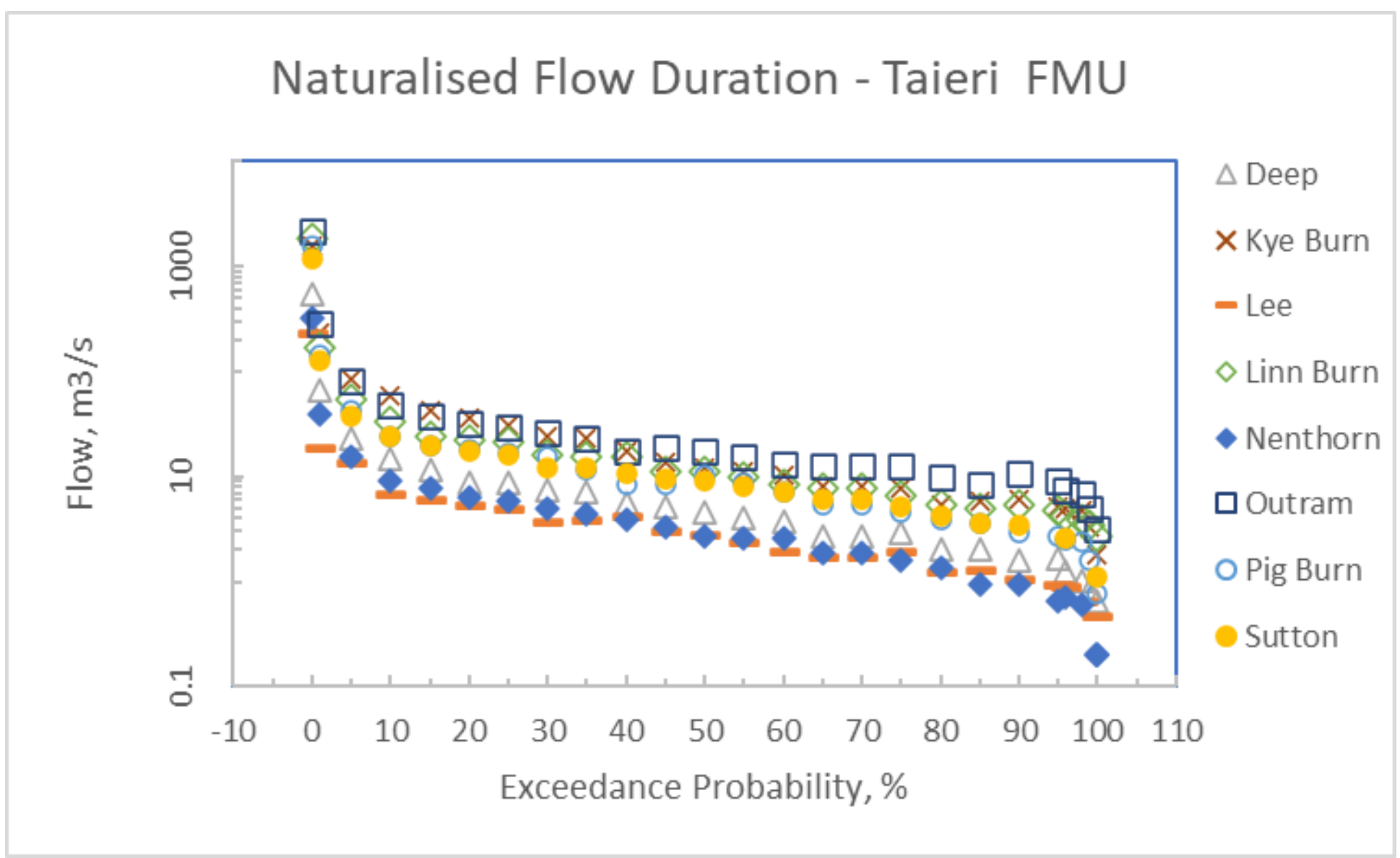

**Fig. 13**. Comparison of median (50$^{th}$ percentile) naturalised flow duration curves among selected gauge sites in the Taieri Freshwater Management Unit, Otago, NZ. Gauge sites are arranged alphabetically: Deep (Deep Stream at SH87) = 411 km$^2$, Kye Burn (Kye Burn Stream at SH85) = 376 km$^2$, Lee (Lee Stream at SH87) = 300 km$^2$, Linn Burn (Taieri at Linn Burn) = 660 km$^2$, Nenthorn (Nenthorn at Mt Stoker Road) = 217 km$^2$. Outram (Taieri at Outram) = 4705 km$^2$. Pig Burn (Pig Burn Stream) = 327 km$^2$, Sutton (Taieri at Sutton Creek) = 3066 km$^2$.

**Table 7.** Summary of naturalised flow duration predictions (50$^{th}$ percentile) at three exceedance probabilities (where 0% = high flow, 50% = median flow, 100% = low flow) among selected gauge sites in the Taieri Freshwater Management Unit, Otago, New Zealand.

| | 50th Percentile Predictions | Exceedance Probabilty | | | | |
|---|---|---|---|---|---|---|
| | | 0% | 50% | 100% | | |
| N | Station | Flow (m3/s) | Flow (m3/s) | Flow (m3/s) | Area (km2) | Strahler Stream Order |
| 1 | Taieri at Outram | 2169 | 18 | 3.2 | 4705 | 7 |
| 2 | Taieri at Sutton | 1647 | 11.5 | 2 | 3066 | 6 |
| 3 | Taieri at Linn Burn | 1828 | 12.3 | 2.9 | 660 | 5 |
| 4 | Kye Burn Stream at SH85 | 1563 | 12.2 | 1.7 | 376 | 4 |
| 5 | Pig Burn Stream | 1552 | 10.2 | 0.79 | 327 | 4 |
| 6 | Deep Stream | 540 | 4.1 | 1.5 | 411 | 4 |
| 7 | Lee Stream at SH87 | 237 | 2.8 | 0.46 | 300 | 4 |
| 8 | Nenthorn at Mt Stoker Road | 317 | 2.7 | 0.2 | 217 | 3 |

In this last section, the naturalised FDCs determined using the set of Meta models are compared with those determined using the calibrated SWAT model (Rajanayaka et al. 2023) for identifying changes to the flow regime. In this regard, there are four possible flow regime degradation scenarios relative to the human influenced FDCs (observations). *Firstly*, there is no degradation to the flow regime. Under this null scenario, the predicted and simulated flow duration curves will be identical to the observations. *Secondly*, the flow regime is completely degraded by human influences. Under this scenario, the predicted and the simulated flow duration curves will be of larger magnitude (shifted upward) than the observations. *Thirdly*, different portions of the flow regime are degraded by human influences. Under this scenario, the predicted and simulated flow duration curves will be shifted upward only at those portions of the flow regime that are degraded. *Fourthly*, the flow regime is not degraded, partly degraded, or completely degraded state but the modeled FDCs are smaller in magnitude (shifted downward) than the observations. Under this scenario, those models producing these results can be considered of poor quality.

The comparison of modeled FDCs is undertaken using physical property information from three-gauge sites along the Taieri River in the Taieri FMU: Taieri at Tiroiti (3095 km$^2$), Taieri at Sutton Creek (3066 km$^2$), and Taieri at Outram (4705 km$^2$). A comparison of predicted and simulated FDCs are presented in Figure 14. In this figure, the current flow regime (observations) appear as red squares, the naturalised Meta model predicted flows at the 50$^{th}$ percentile (median) appear as blue dots, and the naturalised SWAT model simulated flows appear as a black line. At the *Taieri River at Tiroiti* site (catchment area of 3095 km2, Strahler order 5; 13-year period of record from 20/08/1992 to 06/07/2005), the naturalised flows predicted using the Meta model and naturalised flows simulated using the SWAT model appear greater than the observations from median to low flows (exceedance probabilities, e.g., ≥ about 50%) indicating human-influenced degradation of the flow regime (Fig. 14a). By contrast, there are conflicting results among the two models when assessing regime change at peak flows. For example, the naturalised Meta model flows exceed the observations for peak flows (small exceedance probabilities, e.g., ≤ 5%) indicating flow regime degradation, whereas the naturalised SWAT model flows underestimates observations at the 100% exceedance probability. This result implies that the SWAT model is not able to properly simulate naturalised peak flow at this site. Specific Meta model flow values are as follows: at the 0% exceedance probability, the observed value of 504 m3/s, Meta model predicts the median value of 1964 m/3 (range: 964 -6999 m3/s); at the 50% exceedance probability, the observed value is 12.5 m3/s and Meta model predicts a median value of 11.6 m/3 (range: 8.9 -14.3 m3/s); and at the 100% exceedance probability, the observed value is 0.65 m3/s and Meta model predicts a median value of 2.0 m/3 (range: 0.99 -3.6 m3/s).

One conclusion is that the Meta model appears better able to simulate the nonuniform degradation at the Taieri River at Tiroiti site.

At the *Taieri River at Sutton Creek* site (catchment area of 3066 $km^2$, Strahler order 6; 35-year period of record from 9-Jul-1986 to 27-Jan-2021), the naturalised flows predicted using the Meta models are of the same character as at the Taieri River at Tiroiti (Fig. 14b). That is, the naturalised flows predicted using the Meta model appear greater than the observations for peak flows (small exceedance probabilities, e.g., ≤ 5%) and lower flows (large exceedance probabilities, e.g., ≥ about 70%) suggesting selective minor degradation of the flow regime. By contrast, the SWAT model simulated naturalised flows appear to under determine the flow observations across the observed duration curve suggesting the SWAT model is not able to adequately simulate naturalised flow duration at this stie. Specific Meta model flow values are as follows: at the 0% exceedance probability, the observed value of 504 m3/s, Meta model predicts the median value of 1647 m/3 (range: 977.6 -12891 m3/s); at the 50% exceedance probability, the observed value is 12.5 m3/s and Meta model predicts a median value of 11.4 m/3 (range: 8.2 -14.3 m3/s); and at the 100% exceedance probability, the observed value is 0.65 m3/s and Meta model predicts a median value of 2.0 m/3 (range: 0.64 -9.2 m3/s). One conclusion is that the Meta models appear better able to simulate the nonuniform degradation at the Taieri River at Sutton Creek site.

At the *Taieri at Outram* site (catchment area of 4705 $km^2$, Strahler order 6, 35-year period of record from 9-Jul-1986 to 27-Jan-2021), the naturalised flows predicted using the set of Meta models appear greater than the observations for peak flows (small exceedance probabilities, e.g., ≤ 5%) and medium low to low flows (large exceedance probabilities, e.g., ≥ about 70%) suggesting selective degradation in the flow regime (Fig. 14c). By contrast, the SWAT model simulated naturalised flows appear to exceed the peak flow observations and under determine the low flow observations highlighting the poor performance of this model at medium to low flow regime. Specific Meta model flow values are as follows: at 0% exceedance probability the observed flow is 1477 m3/s, the Meta model predicts a naturalised median flow value of 2169 m3/s (range: 1397 - 24705 m3/s), and the SWAT model predicts a median value of 950 m3/s; at 50% exceedance probability the observed flow is 19.8 m3/s, the Meta model predicts a median value of 18 m3/s (range: 13-22.6 m3/s), and the SWAT model predicts a value of 19 m3/s; at the 100% exceedance probability the observed flow is 0.99 m3/s, the Meta model predicts a value of 3.2 m3/s (range: 1.6 – 6.2 m3/s), and the SWAT model predicts a value of 0.1 m3/s. One conclusion from this analysis is that the Meta and SWAT models provide reasonable flow predictions at low exceedance values; however, at flows greater than 40% exceedance probability, the SWAT model is not able to predict reasonable flows at this stie. By contrast, the Meta model predicts reasonable results across the flow regime that are useful for identifying human influenced degradation.

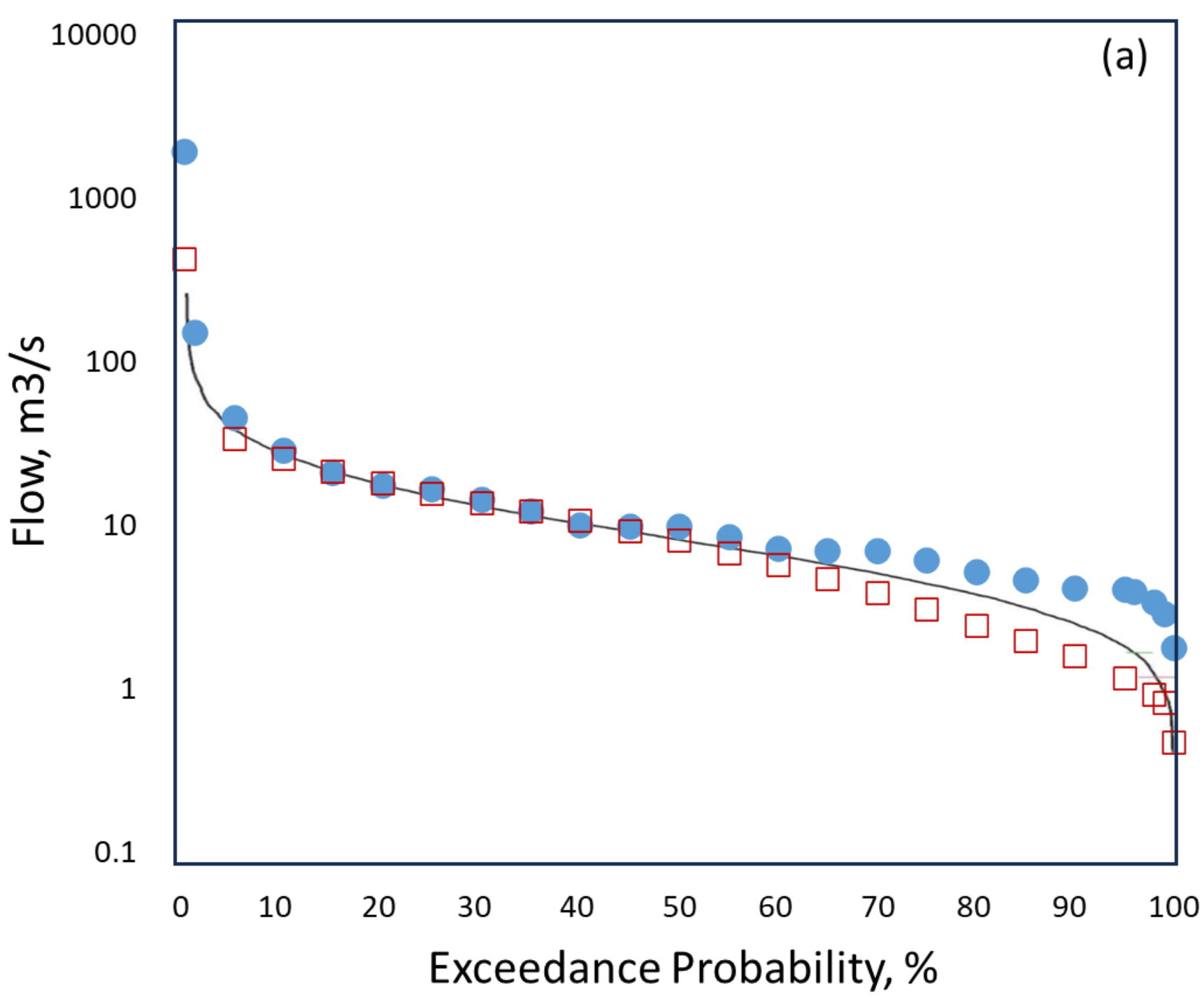

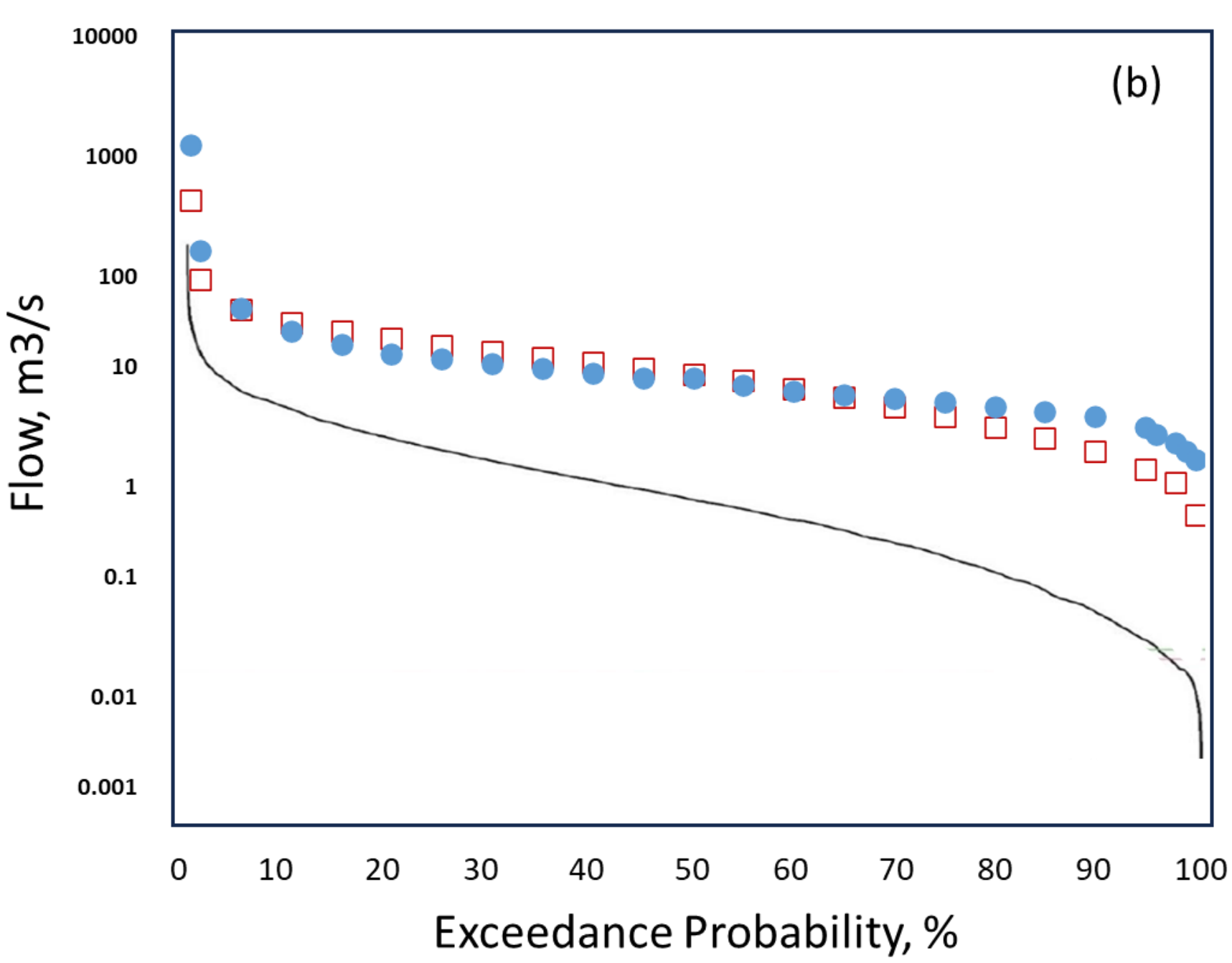

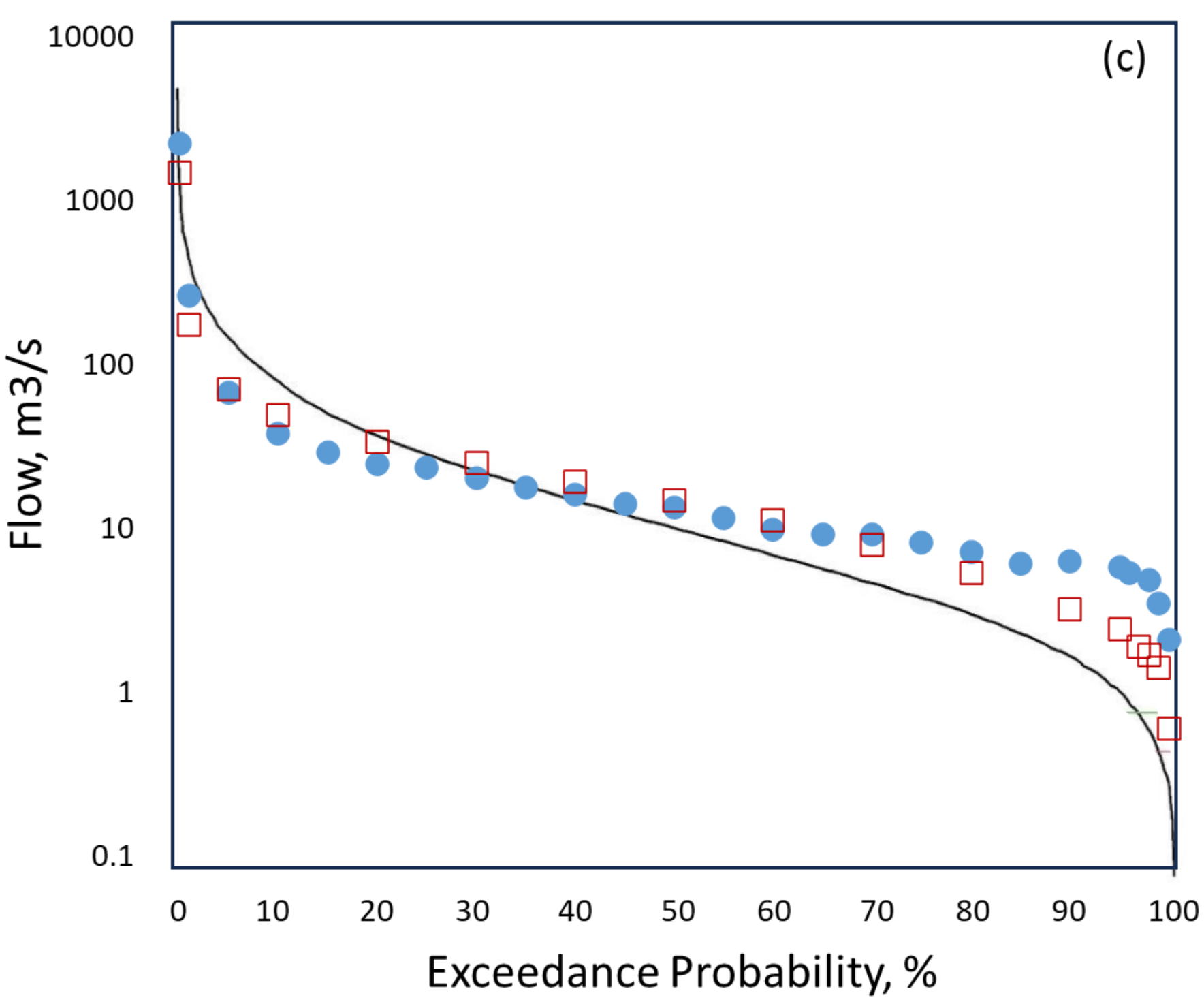

**Fig. 14.** Comparison of naturalised and human influenced flow duration curves for Taieri River gauge sites: (a) The Taieri at Tiroiti (3095 km$^2$), (b) Taieri at Sutton Creek (3066 km$^2$), and (c) Taieri River at Outram (4705 km$^2$). Naturalised Meta model predictions at the 50$^{th}$ percentile appear as filled circles, naturalised deterministic SWAT model simulation appears at a line, and human influenced observations appears as squares.

## 4. Conclusions

This study introduces a new method for predicting naturalized flow duration curves (FDCs) at human influenced sites and multiple catchment scales**.** The combination of natural catchment hydrology and available physical and climate catchment characteristics provide suitable information for Meta model building and prediction of naturalized FDCs and their uncertainty at human influenced catchments. Testing this hypothesis led to the following points.

*First*, a set of Meta models can be successfully developed and used to predict naturalized FDCs and their uncertainty (10$^{th}$, 20$^{th}$, 30$^{th}$, 40$^{th}$, 50$^{th}$, 60$^{th}$, 70$^{th}$, 80$^{th}$, and 90$^{th}$ percentiles) at discrete exceedance probabilities (e.g., 0%, 1%, 3%, 5%, 10%, 15%, 20%, 25%, 30%, 35%, 40%, 45%, 50%, 55%, 60%, 65%, 70%, 75%, 80%, 90%, 95%, 97%, 99%, and 100%) across a range of spatial catchment scales (Strahler order 1 to 7) at human influenced gauged catchments (N=317) and ungauged river reaches (N=18612) across Otago New Zealand.

*Second*, the tradeoff in minimizing base model variance during model development appears to increase the bias in predicted flows below 1 km$^2$ and above 1,000 km$^2$. The application of bias corrections is successfully validated against independent flow observations at the Dart River and Matukituki gauge sites.

*Third*, the extraction of median naturalised flow predictions (50$^{th}$ percentile) at exceedance probabilities that define the FDC facilitates their organization by Strahler stream order. This process demonstrates that the range in flows decrease with increasing exceedance probability (from peak to low flows) and with decreasing Strahler order (large to small catchment areas).

*Fourth*, maps constructed from the 317 gauged and 18612 ungauged reach sites provide visual spatial distribution for naturalised flows at different exceedance probabilities (percent time flows equaled or exceeded) across the Taieri Freshwater Management Unit and adjacent areas, e.g., 0% exceedance probability (percent peak flows equaled or exceeded), 50% exceedance probability (percent median flows equaled or exceeded), and (c) 100% exceedance probability (percent low flows equaled or exceeded).

*Fifth*, the construction of 50$^{th}$ percentile (median) naturalised flow duration curves provides insight on differences in flow regimes at priority gauge sites in the Taieri FMU (ordered from smallest to largest flows): Nenthorn at Mt Stoker Road > Lee at SH87 > Deep Stream at SH87 > Pig Burn Stream > Taieri at Sutton Creek > Kye Burn Stream at SH85, Taieri at Linn Burn, and Taieri at Outram.

*Sixth*, the naturalised FDCs predicted using the Meta models are outperform those simulated using the calibrated SWAT model for the Taieri River gauge sites investigated: Taieri at Tiroiti, Taieri at Sutton Creek, and Taieri River at Outram. Departures in the naturalised reference state are interpreted as flow regime changes across the duration curves.

**Credit authorship contribution statement**

Michael J Friedel: Conceptualization, Methodology, Software, Visualization, Formal analysis, Writing – original draft. Dave Stewart: Data curation, Formal analysis, Investigation, Validation – review & editing. Xaio Feng Lu: Data curation, Formal analysis, Software, Validation, Visualization – review & editing. Pete Stevenson: Data curation, Formal analysis – review & editing. Helen Manly: Resources, Supervision – review & editing; Tom Dyer: Resources and review.

**Declaration of Competing Interest**

The authors declare that they have no known competing financial interests or personal relationships that could have appeared to influence the work reported in this paper.

**Acknowledgement**

This work was supported by the Otago Regional Council, Dunedin, New Zealand [grant number PO033750].